\documentclass[twocolumn]{aastex63}

\usepackage{paralist}
\usepackage{pifont}
\usepackage{bm}
\usepackage{hyperref}

\newcommand{\TRISTAN}{{\bf \textsf{\small Tristan-MP}} }

\renewcommand{\deg}{^{\circ}}

\shorttitle{Return currents \& electron heating}
\shortauthors{Gupta, Caprioli, \& Spitkovsky}
\graphicspath{{./}{figures/}}

\begin{document}

\title{Return Currents in Collisionless Shocks}

\correspondingauthor{Siddhartha Gupta}
\email{gsiddhartha@princeton.edu}

\author[0000-0002-1030-8012]{Siddhartha Gupta}
\affiliation{Department of Astronomy and Astrophysics, The University of Chicago, IL 60637, USA}
\affiliation{Department of Astrophysical Sciences, Princeton University, 4 Ivy Ln., Princeton, NJ 08544, USA}
\author[0000-0003-0939-8775]{Damiano Caprioli}
\affiliation{Department of Astronomy and Astrophysics, The University of Chicago, IL 60637, USA}
\affiliation{Enrico Fermi Institute, The University of Chicago, Chicago, IL 60637, USA}
\author[0000-0001-9179-9054]{Anatoly Spitkovsky}
\affiliation{Department of Astrophysical Sciences, Princeton University, 4 Ivy Ln., Princeton, NJ 08544, USA}

\begin{abstract}
Collisionless shocks tend to send charged particles into the upstream, driving electric currents through the plasma.
Using kinetic particle-in-cell simulations, we investigate how the background thermal plasma neutralizes such currents in the upstream of quasi-parallel non-relativistic electron-proton shocks. 
We observe distinct processes in different regions: the far upstream, the shock precursor, and the shock foot.
In the far upstream, the current is carried by nonthermal protons, which drive electrostatic modes and produce supra-thermal electrons that move towards upstream infinity. 
Closer to the shock (in the precursor), both the current density and the momentum flux of the beam increase, which leads to electromagnetic streaming instabilities that contribute to the thermalization of supra-thermal electrons.
At the shock foot, these electrons are exposed to shock-reflected protons, resulting in a two-stream type instability. 
We analyze these processes and the resulting heating through particle tracking and controlled simulations. 
In particular, we show that the instability at the shock foot can make the effective thermal speed of electrons comparable to the drift speed of the reflected protons.
These findings are important for understanding both the magnetic field amplification and the processes that may lead to the injection of supra-thermal electrons into diffusive shock acceleration.
\end{abstract}
%%%%%%%%%%%%%%%%%%%%%%%%%%%%%%%%%%%%%%%%%%%%%%%%%%%%%%%%%%%
%%%%%%%%%%%%%%%%%%%%%%%%%%%%%%%%%%%%%%%%%%%%%%%%%%%%%%%%%%%
%%%%%%%%%%%%%%%%%%%%%%%  Section 1  %%%%%%%%%%%%%%%%%%%%%%%
%%%%%%%%%%%%%%%%%%%%%%%%%%%%%%%%%%%%%%%%%%%%%%%%%%%%%%%%%%%
%%%%%%%%%%%%%%%%%%%%%%%%%%%%%%%%%%%%%%%%%%%%%%%%%%%%%%%%%%%
\section{Introduction} \label{sec:intro}
%
%%%%%%%%%%%%%%%%%%%%%%%%%%%%%%%%%%%%%%%%%%%%%%%%%%%%%%%%%%%
Energetic charged particles (henceforth cosmic rays -- CRs) are the prime sources of $\gamma$-ray, X-ray, and radio emission across the universe.
Explanation of the acceleration of these particles in shock-powered environments typically relies on the diffusive shock acceleration (DSA) mechanism \citep[e.g.,][]{axford+78,blandford+78, bell78a}, 
where the back-reaction from the streaming supra-thermal (ST) and/or nonthermal (NT) particles on the plasma plays a crucial role \citep[e.g.,][]{bell04,amato+09}.
While the observations of several sources have already provided evidence of self-generated electromagnetic (EM) turbulence in the shock upstream \citep[see e.g.,][]{morlino+10,wilson+16}, whether and how such turbulence helps electrons to be injected into the DSA process remains a pressing unsolved problem.

The properties of the upstream turbulence crucially depend on the ability of the shock to produce back-streaming (i.e., moving toward upstream infinity) ions/protons or electrons, which is controlled by the inclination of the magnetic field relative to the shock normal ($\theta_{\rm Bn}$).
Depending on $\theta_{\rm Bn}$, two main regimes can be identified: (i) quasi-parallel ($\theta_{\rm Bn}\lesssim 50^{\rm o}$) and (ii) oblique/quasi-perpendicular ($\theta_{\rm Bn}\gtrsim 50^{\rm o}$) shocks.
Quasi-parallel shocks reflect protons efficiently \citep[see, e.g.,][]{giacalone+93,caprioli+14a,caprioli+14b,caprioli+15}, producing a current that drives the resonant/nonresonant streaming instabilities \citep[e.g.,][]{bell04,amato+09,bell+13}.
In (quasi-)perpendicular shocks, instead, the current may be carried by electrons \citep[][]{guo+14a,guo+14b,xu+20,kumar+21}, unless the shock magnetization is sufficiently low, in which case ion injection is possible \citep{xu+20,orusa+23}.

In all cases, the question arises of how the CR currents are neutralized.
Current compensation can operate either by the relative drift between thermal protons and electrons in the background plasma or by a set of electrons that travel alongside or in the opposite direction to the current-driving particles, depending on their charge's sign.
These assumptions are usually made in any linear theory of streaming instabilities \cite[see e.g.,][]{amato+09,gupta+21}.
In general, we can expect that if the electrons react to compensate the imbalance, then they are likely ST, because it is hard to envision different populations of thermal electrons with different drifts.
But \emph{are these ST electrons produced locally, or are they shock-reflected energetic electrons?}

While several recent studies have used kinetic simulations to unravel the processes responsible for electron acceleration at shocks \citep[e.g.,][]{sironi+13,guo+14a,park+15,crumley+19,bohdan+19a,xu+20,arbutina+21,kumar+21,shalaby+22,morris+22},
the characterization of return-current electrons has received much less attention, though the generation of ST electrons may be crucial to electron injection, too.

In this paper, we use kinetic particle-in-cell (PIC) simulations to self-consistently study the origin and compensation of currents produced in the shock upstream.
We find that current neutralization gives rise to populations of back-streaming ST electrons, which may be then thermalized closer to the shock; 
the net result is that the electrons impinging on the shock can be much warmer than those far upstream.
We outline our shock simulations in \S \ref{sec:simsetup} and present our main results in \S \ref{sec:shock} and \S \ref{sec:mechanism}.
In \S \ref{sec:fate} and \S \ref{sec:eheating} we design controlled simulations to detail the processes found in global shock simulation and show that the electron heating in the shock foot is universal.
We extend our discussion to quasi-perpendicular shocks in \S \ref{sec:discuss} and summarize our results in \S \ref{sec:conclusion}.
%%%%%%%%%%%%%%%%%%%%%%%%%%%%%%%%%%%%%%%%%%%%%%%%%%%%%%%%%%%
%%%%%%%%%%%%%%%%%%%%%%%%%%%%%%%%%%%%%%%%%%%%%%%%%%%%%%%%%%%
%%%%%%%%%%%%%%%%%%%%%%%  Section 2  %%%%%%%%%%%%%%%%%%%%%%%
%%%%%%%%%%%%%%%%%%%%%%%%%%%%%%%%%%%%%%%%%%%%%%%%%%%%%%%%%%%
%%%%%%%%%%%%%%%%%%%%%%%%%%%%%%%%%%%%%%%%%%%%%%%%%%%%%%%%%%%
\section{Numerical setup}\label{sec:simsetup}
%
%%%%%%%%%%%%%%%%%%%%%%%%%%%%%%%%%%%%%%%%%%%%%%%%%%%%%%%%%%%
We use the EM PIC code \TRISTAN  \citep[][]{spitkovsky05} to perform simulations of collisionless shocks.
The computational domain is quasi-onedimensional, with $5$ cells in the transverse direction ($y$-axis).
Each cell is initialized with an electron-proton plasma, with $200$ particles per cell per species; 
both species have a Maxwell--Boltzmann distribution with a temperature $T_{\rm i,0}=T_{\rm e,0}$ and thermal speeds $v_{\rm th i,e}=\sqrt{k_{\rm B} T_{\rm i,e,0}/m_{\rm i,e}}$ (for ions/protons and electrons, respectively).
Grid spacing and time-steps are fixed to $\Delta x = d_{\rm e}/10$ and $\Delta t = 0.045\, \omega_{\rm pe}^{-1}$, where $\omega_{\rm pe}=\sqrt{4\pi n\,e^2/m_{\rm e}}$ is the electron plasma frequency and  $d_{\rm e}=c/\omega_{\rm pe}$ is the electron skin depth.
The proton skin depth is thus $d_{\rm i}=\sqrt{m_{\rm R}}\,d_{\rm e}$, where $m_{\rm R} = m_{\rm i}/m_{\rm e}$ is the proton-to-electron mass ratio.
As typical in PIC simulations, to save computational resources we use a reduced mass ratio, $m_{\rm R}=100$, though we study how our results depend on $m_{\rm R}$ using controlled simulations.

The magnetization of the plasma is defined by the Alfv\'{e}n speed $v_{\rm A}=B_{\rm 0}/\sqrt{4\pi m_{\rm i} n_{\rm 0}}$; 
the initial magnetic field ${\bf B}_{\rm 0}$ forms an angle $\theta_{\rm Bn}=\cos^{-1}(B_{\rm x}/|{\bf B_0}|)=30\deg$ with the shock normal and lays in the $x-y$ plane (quasi-parallel shock).  
The shock is launched using the left boundary of the computational box as a moving reflecting wall.
Such a piston moves with velocity $v_{\rm pt} \gg v_{\rm A}, v_{\rm th}$ with respect to the background thermal plasma; thus, simulations are in the upstream frame.
Finally, the right boundary is open and expands with time, ensuring the box size remains the minimal necessary to account for back-streaming high-energy particles.

The shock strength is defined by the Alfv\'{e}nic Mach number $\mathcal{M}_{\rm A}\equiv v_{\rm sh}/v_{\rm A}$, and the sonic Mach number $\mathcal{M}_{\rm s}\equiv v_{\rm sh}/v_{\rm th,i}$, where $v_{\rm sh}=v_{\rm pt}\mathcal{R}/(\mathcal{R}-1)$ and $\mathcal{R}\equiv\rho_{\rm 2}/\rho_{\rm 1}$ is the shock  compression.
To obtain the shock parameters, we have assumed $\mathcal{R}=4$, as is typical for strong shocks, though the actual value of $\mathcal{R}$ may be larger than $4$ \citep{haggerty+20,caprioli+20}.
Finally, we introduce $\tilde{p}\equiv p/m_i v_{\rm pt}$ as the particle momentum normalized to the momentum of a proton moving with the piston speed (henceforth piston momentum). 
%%%%%%%%%%%%%%%%%%%%%%%%%%%%%%%%%%%%%%%%%%%%%%%%%%%%%%%%%%%
%%%%%%%%%%%%%%%%%%%%%%%%%%%%%%%%%%%%%%%%%%%%%%%%%%%%%%%%%%%
%%%%%%%%%%%%%%%%%%%%%%%  Section 3  %%%%%%%%%%%%%%%%%%%%%%%
%%%%%%%%%%%%%%%%%%%%%%%%%%%%%%%%%%%%%%%%%%%%%%%%%%%%%%%%%%%
%%%%%%%%%%%%%%%%%%%%%%%%%%%%%%%%%%%%%%%%%%%%%%%%%%%%%%%%%%%
\section{The Nature of the Return Current} \label{sec:shock}
%
%%%%%%%%%%%%%%%%%%%%%%%%%%%%%%%%%%%%%%%%%%%%%%%%%%%%%%%%%%%
\begin{figure*}
\centering
    \includegraphics[width=7.in]{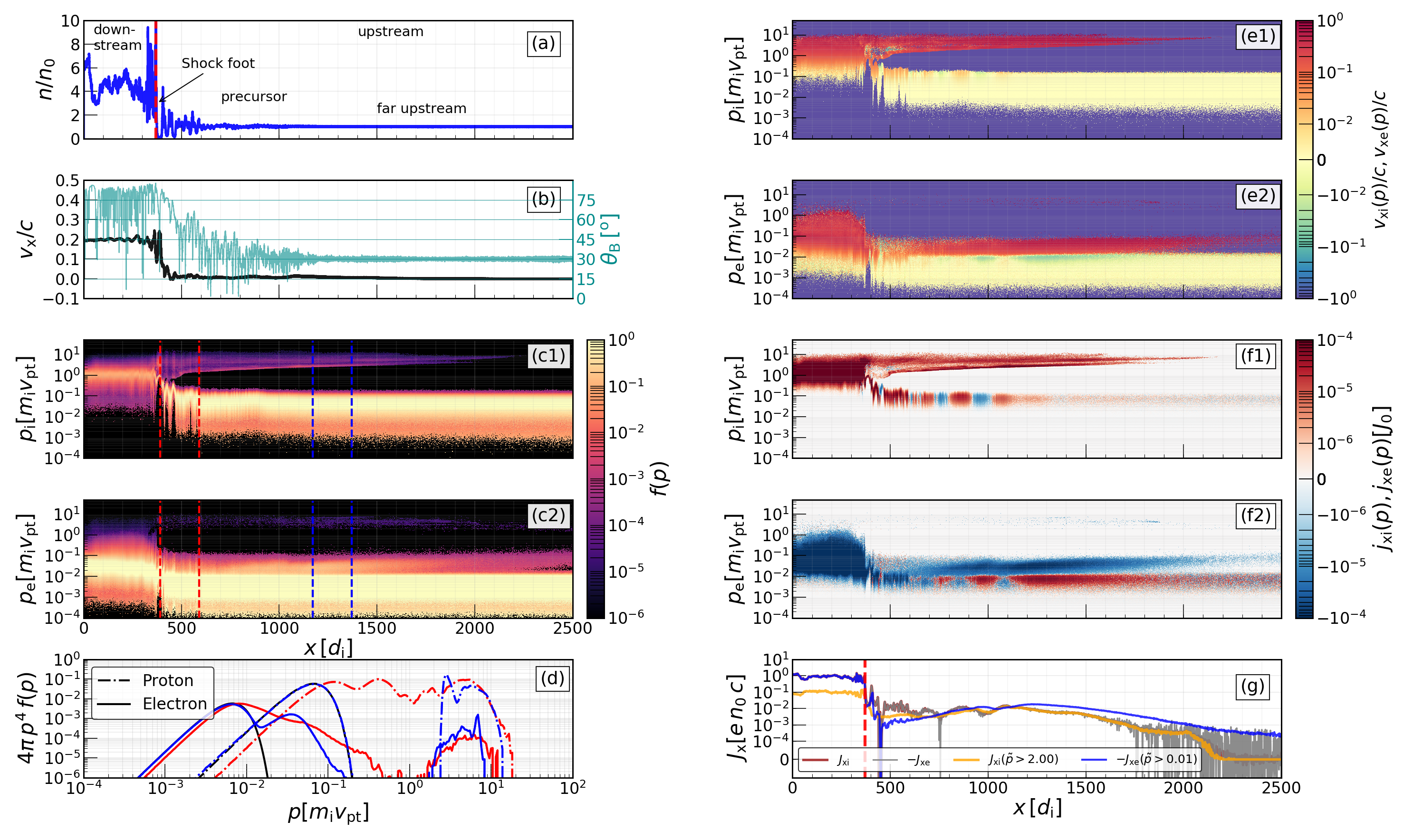}
    \caption{
    Profiles of different physical quantities at $t=85\,\omega_{\rm ci}^{-1}$ for our benchmark shock simulation with $v_{\rm pt}/c=0.2$, $\mathcal{M}_{\rm A}=20$, $\mathcal{M}_{\rm s}=40$, $\theta_{\rm Bn}=30\deg$, and $m_{\rm R}=100$.
    Panels (a) and (b) display density and total (thermal + CRs) $x$-velocity.
    Panels (c1) and (c2) show the $x-p$ phase-space for protons and electrons, and panel (d) shows their momentum spectra in the two upstream regions marked in panels (c1, c2).
    The $x$-velocity (Equation \ref{eq:f_vp}) and current density (Equation \ref{eq:f_jp}) as a function of momentum $p$ are displayed in panels (e1, e2) and (f1, f2) respectively. 
    Panel (g) shows total and partial currents, as in the legend; note that although the net currents are balanced, NT protons ($p/m_{\rm i}v_{\rm pt}\gtrsim 2$, orange) and ST+NT electrons ($p/m_{\rm i}v_{\rm pt}\gtrsim 0.01$, blue) constitute currents that are sometimes larger or smaller than the total current in the respective species.
}
    \label{fig:keyfig}
\end{figure*}
Figure \ref{fig:keyfig} illustrates the main properties of our benchmark shock, with $v_{\rm pt}=0.2\, c, \mathcal{M}_{\rm A}=20, \mathcal{M}_{\rm s}=40$.
Figure \ref{fig:keyfig}(a) shows the plasma density profile, with the red line marking the shock location at $t=85\,\omega_{\rm ci}^{-1}$;
the region right (left) of this line represents upstream (downstream).
Figure \ref{fig:keyfig}(b) displays the profile of the $x$-component of the plasma velocity (left axis, black curve) and the local inclination of the magnetic field relative to the shock normal (right axis, cyan).
Close to the shock $\theta_{\rm Bn}$ can deviate significantly from $30\deg$, due to the magnetic field amplification driven by the current in back-streaming energetic protons \citep[e.g.,][]{bell04,riquelme+09,caprioli+14b,park+15}.

Figures \ref{fig:keyfig}(c1, c2) show the phase-space distribution $f(x,p)$ of protons and electrons, respectively; their spectra in two regions upstream (marked with red and blue lines in Figures \ref{fig:keyfig}(c1,c2)) are reported in Figure \ref{fig:keyfig}(d).
Figures \ref{fig:keyfig}(e1, e2) show the $x$-component of the particle velocity, $v_{\rm x}(x,p)$, and Figures \ref{fig:keyfig}(f1, f2) illustrate the specific current $j_{\rm x}(x,p)$ carried by protons and electrons with momentum $p$, which are defined as
\begin{eqnarray}
\label{eq:f_vp}
v_{\rm x}(x,p)&\equiv &\frac{\int_{p}^{p+\Delta p} u_{\rm x}(x,p')f(x,p')p'^2 dp'}{\int_{p}^{p+\Delta p} f(x,p')p'^2 dp'}
\end{eqnarray}
\begin{eqnarray}
\label{eq:f_jp}
j_{\rm x}(x,p)&=&\int_{p}^{p+\Delta p} q \,u_{\rm x}(x,p')f(x,p')p'^2 dp'.
\end{eqnarray}
Here the total momentum $p'$ is measured in the upstream frame, $u_{\rm x}$ is the $x$-component of particle's $3$-velocity, $q=\pm e$ (depending on the species), and $\Delta p$ is an infinitesimal momentum increment used for estimating distributions.
The vertical axes in Figures \ref{fig:keyfig}(c1, c2), Figures \ref{fig:keyfig}(e1, e2), and Figures \ref{fig:keyfig}(f1, f2) are normalized to the piston momentum.

{\it Momentum distributions}: In Figures \ref{fig:keyfig}(c1,c2) the golden-yellow regions ahead of the shock correspond to the thermal populations.
The proton phase-space (Figure \ref{fig:keyfig}(c1)) contains a population of high-momentum particles ($\tilde{p}\equiv p/m_{\rm i}v_{\rm pt}\gtrsim 2$) undergoing DSA (hereafter NT, or DSA, protons).
The electron phase-space (Figure \ref{fig:keyfig}(c2)), besides such populations, also contains a third, intermediate, one: electrons with $\tilde{p}\sim 0.01$ (i.e., above their thermal distributions), which we label as supra-thermal (ST); 
the origin of such ST electrons is investigated in this paper.

{\it Velocity distributions}: The $x-p$ phase-space distribution of $v_x$ (Equation \ref{eq:f_vp}) for each species is shown in Figures \ref{fig:keyfig}(e1,e2).
While far upstream thermal protons and electrons are at rest, close to the shock they acquire finite velocities to compensate the currents of back-streaming energetic particles (both NT protons and ST electrons).
%=======================================
\begin{figure*}[ht!]
\begin{minipage}{.29\linewidth}
\centering
\includegraphics[height=3.05in]{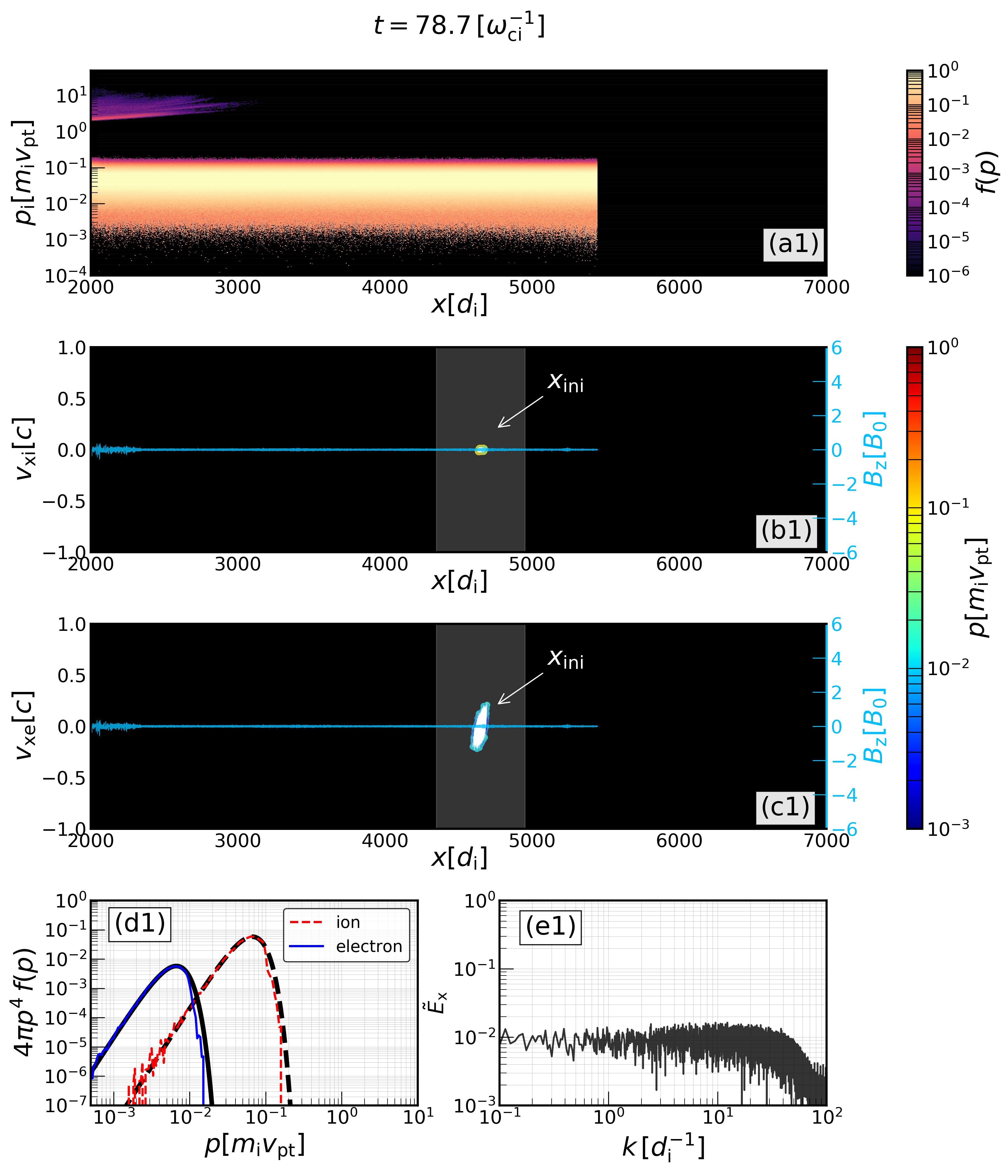}
\end{minipage}
\hspace{0.02\linewidth}
\begin{minipage}{.29\linewidth}
\centering
\includegraphics[height=3.05in]{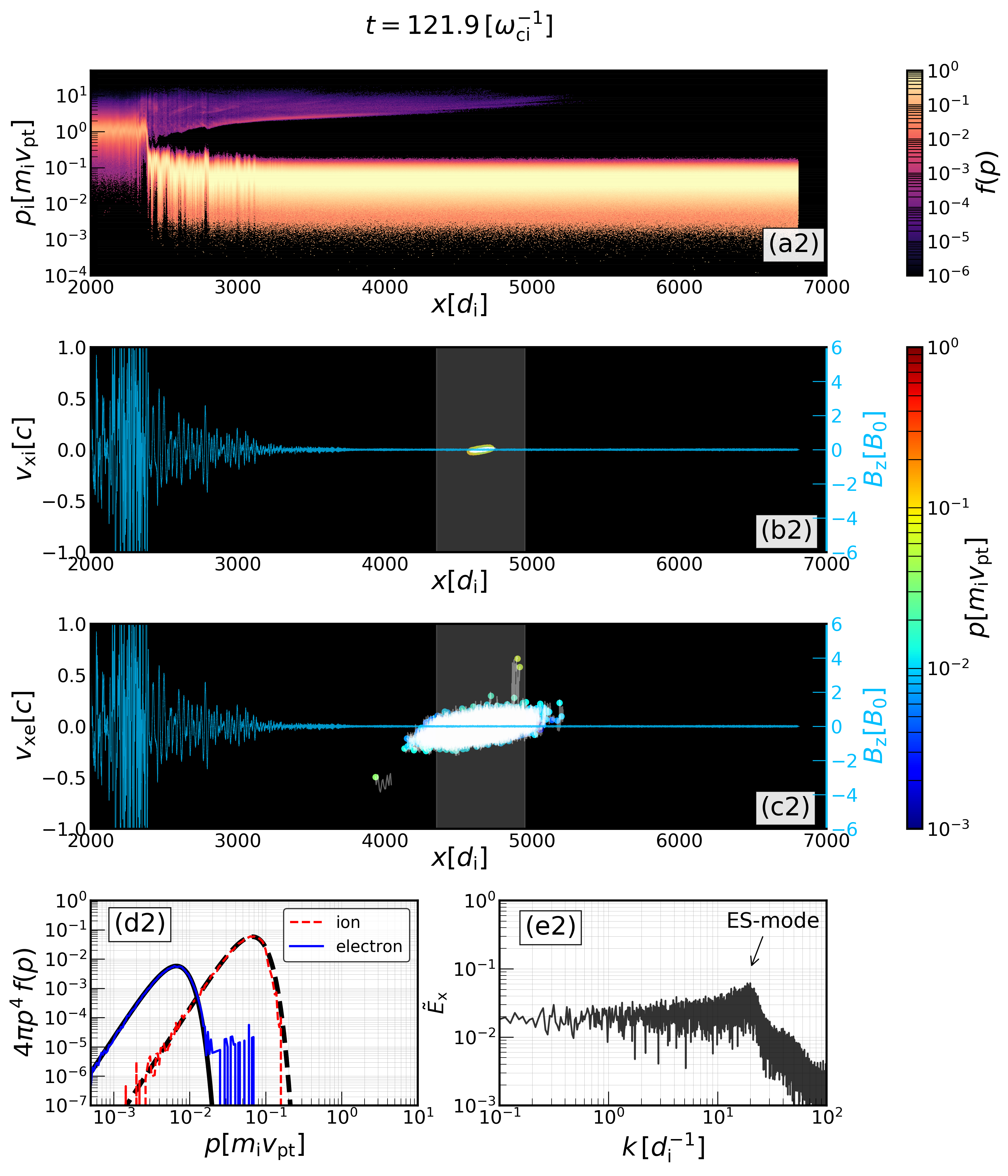}
\end{minipage}
\hspace{0.02\linewidth}
\begin{minipage}{.31\linewidth}
\centering
\includegraphics[height=3.05in]{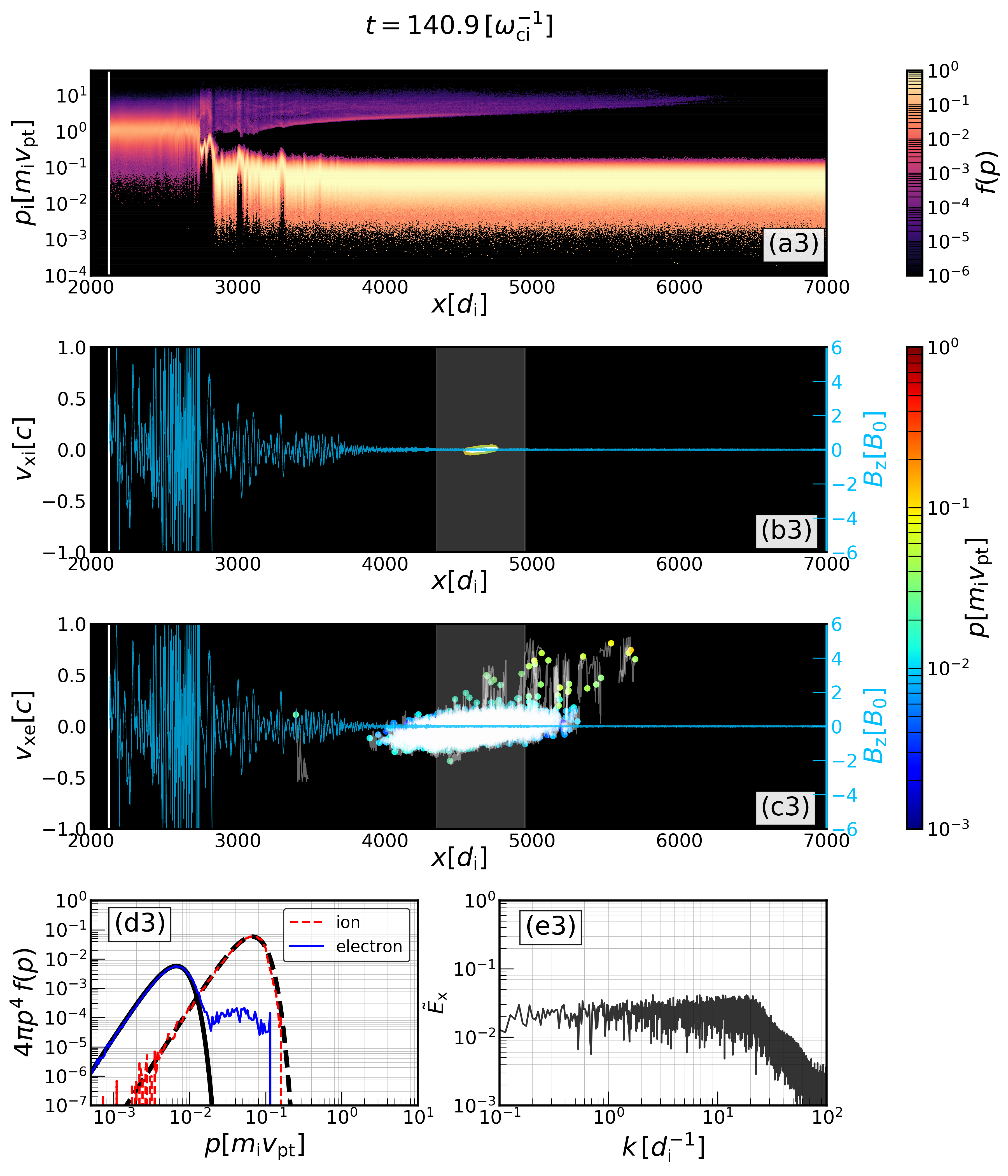}
\end{minipage}
\caption{Development of the ST electrons in the far upstream of a quasi-parallel shock.
Columns from left to right correspond to $t=78.7\,\omega_{\rm ci}^{-1}$, $121.9\,\omega_{\rm ci}^{-1}$, and $140.9\,\omega_{\rm ci}^{-1}$.
Panels (a1--a3) show the proton $x-p$ phase-space, while panels (b1--b3,c1--c3) show the $x-v_{\rm x}$ phase-space of tracer protons and electrons; the blue curves indicate $B_{\rm z}$. 
Note that the horizontal axes in these panels represent the distance in the upstream rest frame.
Panels (d1--d3) show the momentum distribution of the tracked electrons/protons and panels (e1--e3) display the mode analysis of $|E_{\rm x}|$ in the highlighted region of panels (b1--b3 or c1--c3);
when NT particles enter the region (middle column), protons remain thermal, however electrons develop a ST, back-streaming population well before they reach the shock.}
\label{fig:tracrealparticle}
\end{figure*}
%=======================================

{\it Current distributions}: Figures \ref{fig:keyfig}(f1, f2) show the currents carried by particles with momentum $p$ (Equation \ref{eq:f_jp}), while Figure \ref{fig:keyfig}(g) shows the total currents in protons and electrons (brown and grey curves), integrated above $\tilde{p}=2$ for NT protons (orange) and above $\tilde{p}=0.01$ for ST+NT electrons (blue).
Close to the shock, the NT proton current $j_{\rm xi}$ is comparable to the ST electron current.
This is a non-trivial finding: in the literature, the proton-driven current is usually assumed to be compensated by thermal electrons -- a condition that is imposed by hand \citep{bell04,amato+09}; 
here, instead, we find that the shock naturally develops a population of back-streaming energetic electrons, whose density and drift speed are regulated to compensate the proton current.
At a sufficiently large distance upstream, though, where EM turbulence is weak or has not grown yet, the ST electrons travel almost freely and typically faster than NT protons.
This implies that upstream of the back-streaming protons, there must be another, transient, current imbalance, this time driven by escaping ST electrons, which is compensated by the background thermal electrons.

The discussion above shows that the particle distributions upstream of quasi-parallel shocks readjust automatically in response to the currents launched by the shock. 
In \S\ref{sec:mechanism} -- \S\ref{sec:eheating} we quantify how and where the ST electrons are produced, and the ensuing electron heating.
%%%%%%%%%%%%%%%%%%%%%%%%%%%%%%%%%%%%%%%%%%%%%%%%%%%%%%%%%%%
%%%%%%%%%%%%%%%%%%%%%%%%%%%%%%%%%%%%%%%%%%%%%%%%%%%%%%%%%%%
%%%%%%%%%%%%%%%%%%%%%%%  Section 4  %%%%%%%%%%%%%%%%%%%%%%%
%%%%%%%%%%%%%%%%%%%%%%%%%%%%%%%%%%%%%%%%%%%%%%%%%%%%%%%%%%%
%%%%%%%%%%%%%%%%%%%%%%%%%%%%%%%%%%%%%%%%%%%%%%%%%%%%%%%%%%%
\section{Production sites and mechanisms}\label{sec:mechanism}
%
%%%%%%%%%%%%%%%%%%%%%%%%%%%%%%%%%%%%%%%%%%%%%%%%%%%%%%%%%%%
In this section we study the microphysics of generation of ST electrons that provide the return current. 
We focus on two distinct regions: far upstream and close to the shock, as illustrated in the following.
%===============================================================
\subsection{Far Upstream} \label{subsec:far-up}
%
%===============================================================
Let us consider a region far from the shock and investigate the response of background plasma to NT protons. 
To do this, we track $(4+4)\times 10^{4}$ thermal protons and electrons starting in the pristine upstream medium (position $x_{\rm ini}$); 
their evolution at $t= 78.7\,\omega_{\rm ci}^{-1}$, $121.9\,\omega_{\rm ci}^{-1}$, and $140.9\,\omega_{\rm ci}^{-1}$ is shown in three columns of Figure \ref{fig:tracrealparticle}.

Figures \ref{fig:tracrealparticle}(a1--a3) show the proton $x-p$ phase-space.
The $x-v_{\rm x}$ phase-space of the tracked protons and electrons are shown in Figures \ref{fig:tracrealparticle}(b1--b3, c1--c3), along with the profile of $B_{\rm z}$, with particles color-coded according to their momentum (rightmost color bar).
Figures \ref{fig:tracrealparticle}(d1--d3) show the spectrum of the tracked protons and electrons, which are initially thermal as seen by comparing the blue and red curves with black (Maxwellian) curves.
Figures \ref{fig:tracrealparticle}(e1--e3) show the Fourier analysis of $E_{\rm x}$ in the grey region of Figures \ref{fig:tracrealparticle}(b1--b3, c1--c3).

When the NT protons reach $x_{\rm ini}$ (the middle column), a distinct electrostatic (ES) mode appears (Figure \ref{fig:tracrealparticle}(e2)), which is due to interactions between the current beam and thermal plasma.
The wavenumber of the mode is found to be (see Appendix \ref{app:dispersion})
\begin{equation}\label{eq:kfast}
k = \frac{\omega_{\rm pe}}{v_{\rm ib}} = \frac{\sqrt{m_{\rm R}}}{v_{\rm ib}/c}\,d_{\rm i}^{-1}\,
,
\end{equation}
where $v_{\rm ib}\gtrsim 2\,v_{\rm pt}$ is the drift velocity of the current driving particles.
The middle column also shows that the tracked protons remain thermal, while several electrons achieve ST momenta and start moving towards upstream infinity.
The energization of these electrons is evident in Figure \ref{fig:tracrealparticle}(d3), corresponding to $t=140.9\,\omega_{\rm ci}^{-1}$, where the electron distribution exhibits a ST hump similar to Figure \ref{fig:keyfig}(d).
Note the ES mode has also disappeared, which means that the proton beam has eventually transferred a fraction of its energy to the background electrons.

As the shock moves, the NT beam becomes less anisotropic and the current in the region $x_{\rm ini}$ increases, which can drive EM instabilities; the whole background will eventually thermalize to a higher temperature. 
However, the above scenario may still be valid in the far upstream where the charged beam penetrates the fresh plasma, as long as the current in the beam is not negligible.
Thus the result provides clear evidence that: 1) the ST electrons that contribute to the return current can be produced locally in the upstream, and 2) a population of back-streaming electrons in the upstream is not necessarily a signature of injection into DSA.
\begin{figure}[t!]
\includegraphics[width=3.5in]{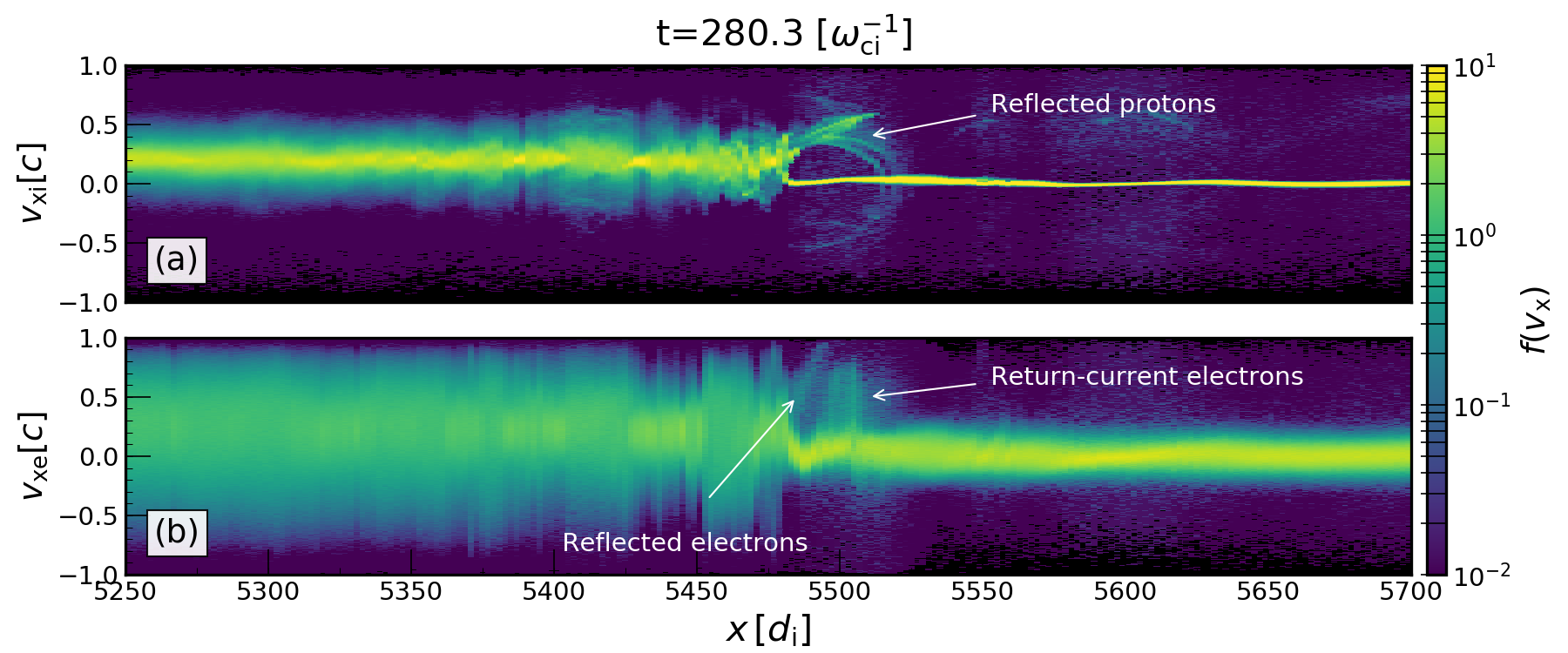}
\includegraphics[width=3.5in]{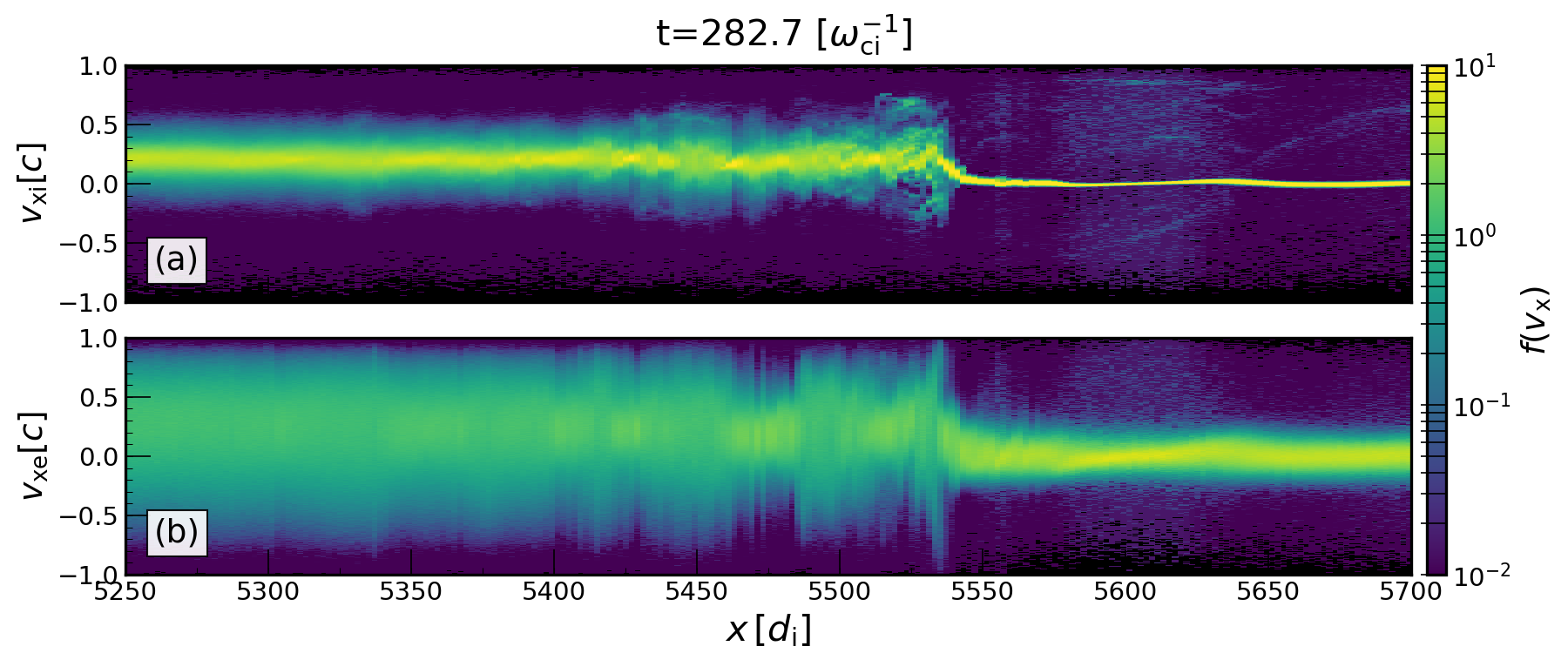}
\caption{
$x-v_{\rm x}$ phase-space distribution of protons (panel a) and electrons (panel b) at $t=280.3\,\omega_{\rm ci}^{-1}$ and $282.7\,\omega_{\rm ci}^{-1}$ (top and bottom panels, respectively).
The arrows in the top panels indicate the specularly-reflected protons and the two distinct populations of high-$v_x$ electrons: those reflected at the shock ramp and those produced by the proton current. 
Both reflected populations vanish in the bottom panels.}
\label{fig:timesnap_x_vx}
\end{figure}
%===============================================================
\subsection{Immediately Upstream of the Shock}\label{subsec:foot}
%
%===============================================================
The characterization of the return current very close to the shock is challenging, since different instabilities operate simultaneously.
On top of the streaming instability driven by DSA protons, the shock foot (i.e., the region immediately upstream of the shock) also experiences the current of specularly reflected thermal protons, which drives the reformation of any super-critical shock \citep[e.g.,][]{treumann09} on a cyclotron timescale \citep[e.g.,][]{thomas+90,lee+04,caprioli+15}.
These reflected protons can create a scenario similar to the far upstream (\S \ref{subsec:far-up}), as we illustrate below.

\begin{figure}[t!]
\includegraphics[width=3.5in]{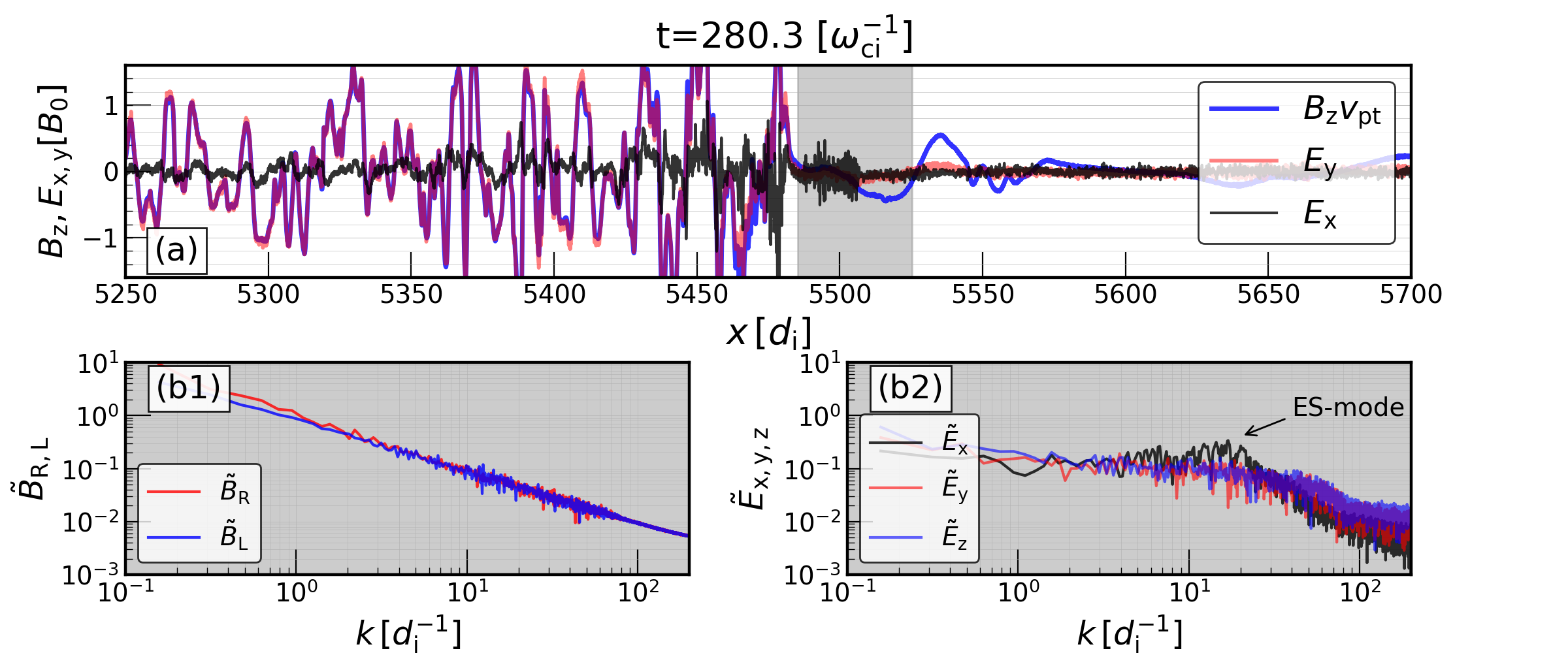}
\includegraphics[width=3.5in]{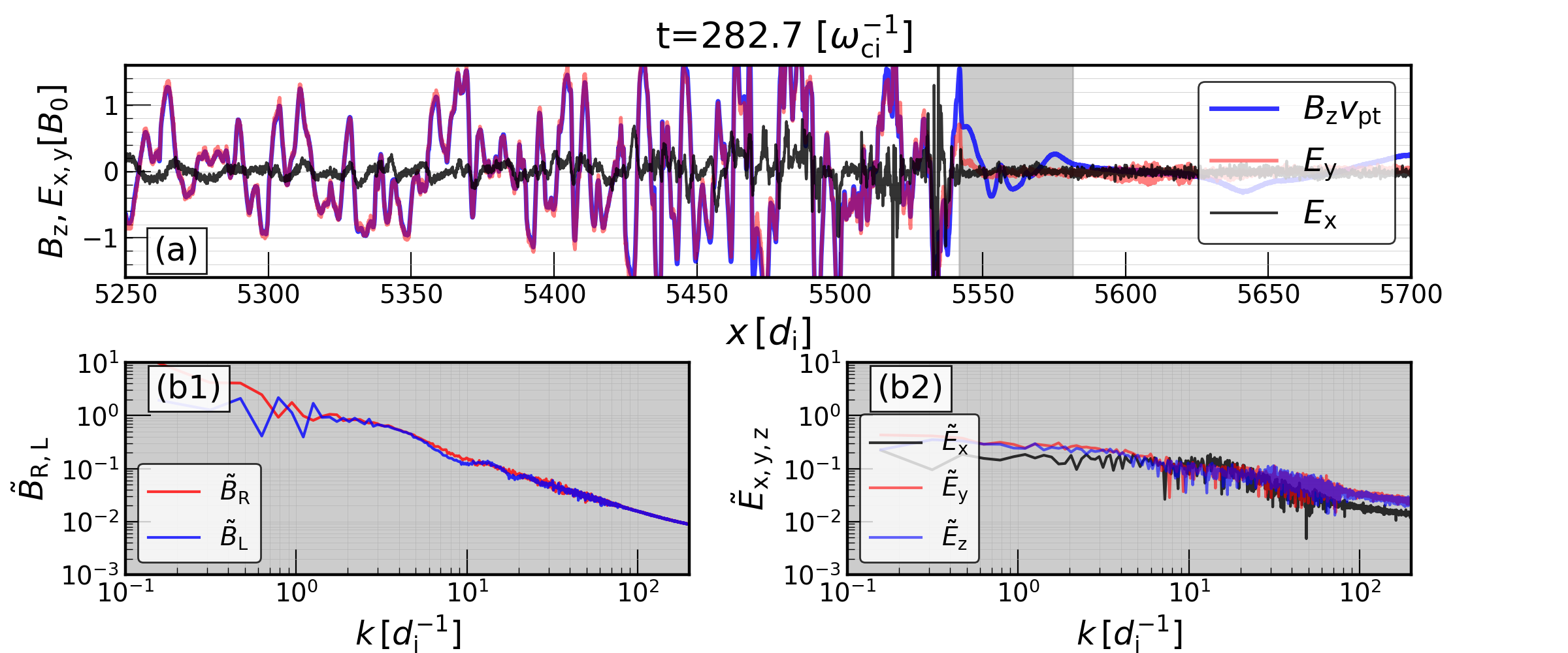}
\caption{Diagnostics of the EM profiles at two different times as shown in Figure \ref{fig:timesnap_x_vx}.
The panel (a) shows the profiles of $E_{\rm x}, E_{\rm y},$ and $B_{\rm z}$.
The foot of the shock is highlighted by the grey region.
The panels (b1) show the Fourier analysis of the right and left circular modes ($\tilde{B}_{\rm R,L}=\tilde{B}_{\rm y}\pm j \tilde{B}_{\rm z}$) and panel (b2) represents the Fourier analysis of $E_{\rm x},E_{\rm y},$ and $E_{\rm z}$ in the grey region.
When the shock is reforming (at $t=280.3\,\omega_{\rm ci}^{-1}$), the appearance of an ES mode is evident.
}
\label{fig:timesnap_fld}
\end{figure}
Figures \ref{fig:timesnap_x_vx}(a) and \ref{fig:timesnap_x_vx}(b) show the evolution of the $x-v_{\rm x}$ phase-space for protons and electrons between $t=280.3\, \omega_{\rm ci}^{-1}$ and $282.7\,\omega_{\rm ci}^{-1}$.
Times are chosen in order to illustrate one cycle of shock reformation.
Initially, we see a beam of specularly-reflected protons with velocity $v_{\rm ib}\approx 0.4 c$ ($\equiv 2 v_{\rm pt}$);
the left edge of the proton beam represents the old location of the shock, and the right one marks the location where the new shock barrier will reform.
At $t=280.3\,\omega_{\rm ci}^{-1}$ the electron phase-space shows two distinct populations: one close to the old shock barrier and another in the middle of the proton beam, indicated by the arrows.
The former population corresponds to shock-reflected electrons, while the latter arises directly because of the proton beam and not because of reflection from the shock;
the electrons in this ST population have a $v_x$ velocity that exceeds the shock speed, which means that they stream away upstream.
Both features fade away when the proton beam disappears (i.e., when the new barrier is formed), as shown in the snapshot at $282.7\,\omega_{\rm ci}^{-1}$.
Thus, the shock reformation periodically produces a population of back-streaming electrons.

To understand how the back-streaming electrons formed, we plot the profiles of $E_{\rm x,y}$ and $B_{\rm z}$ in Figure \ref{fig:timesnap_fld}(a) and their mode diagnostics in Figures \ref{fig:timesnap_fld}(b1,b2), for the same two epochs considered in Figure \ref{fig:timesnap_x_vx}.
First, we notice that in the downstream $E_{\rm y}\approx v_{\rm pt} B_{\rm z}/c$, as expected for the transverse motional electric field ($y-z$) produced due to the bulk displacement of downstream plasma in the upstream frame. The parallel component $E_{\rm x}$ cannot be totally motional and in fact develops due to the pressure anisotropy in different populations.
Importantly, the Fourier analysis shows that a distinct ES mode appears around $k\approx 20\,d_{\rm i}^{-1}$ at $280.3\,\omega_{\rm ci}^{-1}$, i.e., when the proton current is strong (Figure \ref{fig:timesnap_x_vx}). 
The wavenumber of this mode is consistent with Equation \ref{eq:kfast}, which indicates that the nature of the instability is similar to the two-stream instability or the Buneman instability \citep[][]{hoshino+02,bret+10b,muschietti+17}.
This instability is modulated with the quasi-periodic proton reflection and produces a parallel $E_{\rm x}$ that can scatter and energize electrons.
%%%%%%%%%%%%%%%%%%%%%%%%%%%%%%%%%%%%%%%%%%%%%%%%%%%%%%%%%%%
%%%%%%%%%%%%%%%%%%%%%%%%%%%%%%%%%%%%%%%%%%%%%%%%%%%%%%%%%%%
%%%%%%%%%%%%%%%%%%%%%%%  Section 5  %%%%%%%%%%%%%%%%%%%%%%%
%%%%%%%%%%%%%%%%%%%%%%%%%%%%%%%%%%%%%%%%%%%%%%%%%%%%%%%%%%%
%%%%%%%%%%%%%%%%%%%%%%%%%%%%%%%%%%%%%%%%%%%%%%%%%%%%%%%%%%%
%%%%%%%%%%%%%%%%%%%%%%%%%%%%%%%%%%%%%%%%%%%%%%%%%%%%%%%%%%%
%
\section{The fate of return-current electrons} \label{sec:fate}
%
%%%%%%%%%%%%%%%%%%%%%%%%%%%%%%%%%%%%%%%%%%%%%%%%%%%%%%%%%%%
In the previous section, we have discussed how ST electrons are produced locally at the shock and far upstream in response to the proton-driven current. 
In a global shock simulation, at any point in the upstream, one in principle has the contribution from ST electrons produced in the shock foot, as well as the ones locally produced by the streaming protons and those reflected from the shock.
In order to disentangle such contributions and to assess the long-term evolution of such ST electrons, we perform controlled simulations with different proton beams without including the shock.
In the end, we show that depending on the distribution of the proton beam and the time for which electrons are exposed to it, their evolution can be different.
%===================================================
\subsection{Setup of Controlled Simulations}\label{subsec:setupcontrollsim}
%
%===================================================
We set up the thermal plasma parameters similar to our shock run, but with periodic boundary conditions in all directions.
In addition to background electrons and protons, we introduce a third species that represents the beam protons;
the beam-to-background proton density ratio, $n_{\rm ib}/n_{\rm 0}$, is implemented by varying the weight of the species \citep[see also][]{riquelme+09,gupta+21}.
In the benchmark run, we apply a boost speed of $v_{\rm bst}=0.4\,c$ ($\equiv 2\, v_{\rm pt}$) 
%\dam{seems that we need a name for this boost}
to the current-driving protons, without imposing any bulk speed on the background plasma to compensate for such a current.
In this way, we can observe the development of the return current from scratch.
Nevertheless, we observed that the outcomes remained unaffected, even when the return current was introduced at the initial time ($t=0$).
One remaining parameter is the velocity distribution of the beam protons that determines the drift speed $v_{\rm ib}$ relative to the background plasma frame in this setup \citep[see equation 12 in][]{gupta+21} is chosen as follows.
\begin{table}[t!]
    \centering
    \begin{tabular}{l l l l l }
    \hline
         \hline
          %&  Beam & Momentum  & Density  & Velocity  \\
          Region & Beam type & $n_{\rm ib}/n_{\rm 0}$  & $v_{\rm ib}/c$ & $p_{\rm ib}/m_{\rm i}c$\\
         \hline
        1. Far upstream & dilute cold   & $0.005$ & $0.4$ & $0.4$ \\
        2. Precursor & dilute hot       & $0.005$ & $0.3$ & $0.5$\\
        3. Shock-foot & dense cold      & $0.1$   & $0.4$ & $0.4$ \\
         \hline
    \end{tabular}
    \caption{
    The beam parameters for our benchmark controlled simulations, representing three distinct regions in the upstream, as shown in Figure \ref{fig:keyfig}(a).
    For the background plasma: $v_{\rm A}/c=1.33\times 10^{-2}$, $v_{\rm th}/c=6.67\times 10^{-3}$,
    and $m_{\rm R}=100$.
    The drift velocity and the average momentum along the $x$-direction are provided in the background plasma rest frame (obtained using equations $12$ and $13$ in \citealt{gupta+21}).
    }
    \label{tab:controlsim}
\end{table}

We consider two cases: 1) cold beam with the speed of particles in the beam rest-frame $u_{\rm iso}\simeq 0$ (or $u_{\rm iso}\ll v_{\rm bst}$) and 2) hot beam with $u_{\rm iso}\gtrsim v_{\rm ib}$.
The former case is a representative of the shock-foot (specularly reflected protons, e.g., Figure \ref{fig:timesnap_x_vx}) or far-upstream region (particles escaping upstream infinity, e.g., Figure \ref{fig:tracrealparticle}),
whereas the latter case stands for the shock precursor, where the beam is diffusing.
The run parameters of these simulations are listed in Table \ref{tab:controlsim}.
While such numbers may vary depending on the shock speed or the momentum/energy of the NT particles, the beam type and consequently its influence on the plasma are expected to remain consistent with the findings that we explore in the following sections.
Note that although we discuss dilute/dense cold/hot cases separately, in a realistic scenario the plasma experiences both as the background thermal populations move close to the shock.
%===================================================
\subsection{Results}\label{subsec:controlledsimresult}
%
%===================================================
%===================================================
\subsubsection{Cold Beams}\label{subsubsec:cold}
%
%===================================================
We consider two limiting regimes: a dilute beam ($n_{\rm ib}/n_{\rm 0}\lesssim  0.01$, region $1$ in Table \ref{tab:controlsim}) and a dense beam ($n_{\rm ib}/n_{\rm 0}\sim 0.1$, region $3$).
The dilute beam is representative of the beam of energetic protons streaming away far upstream, while the dense beam in the shock foot is produced by the specular reflection at the shock.
\begin{figure}
%\centering
\includegraphics[width=3.4in]{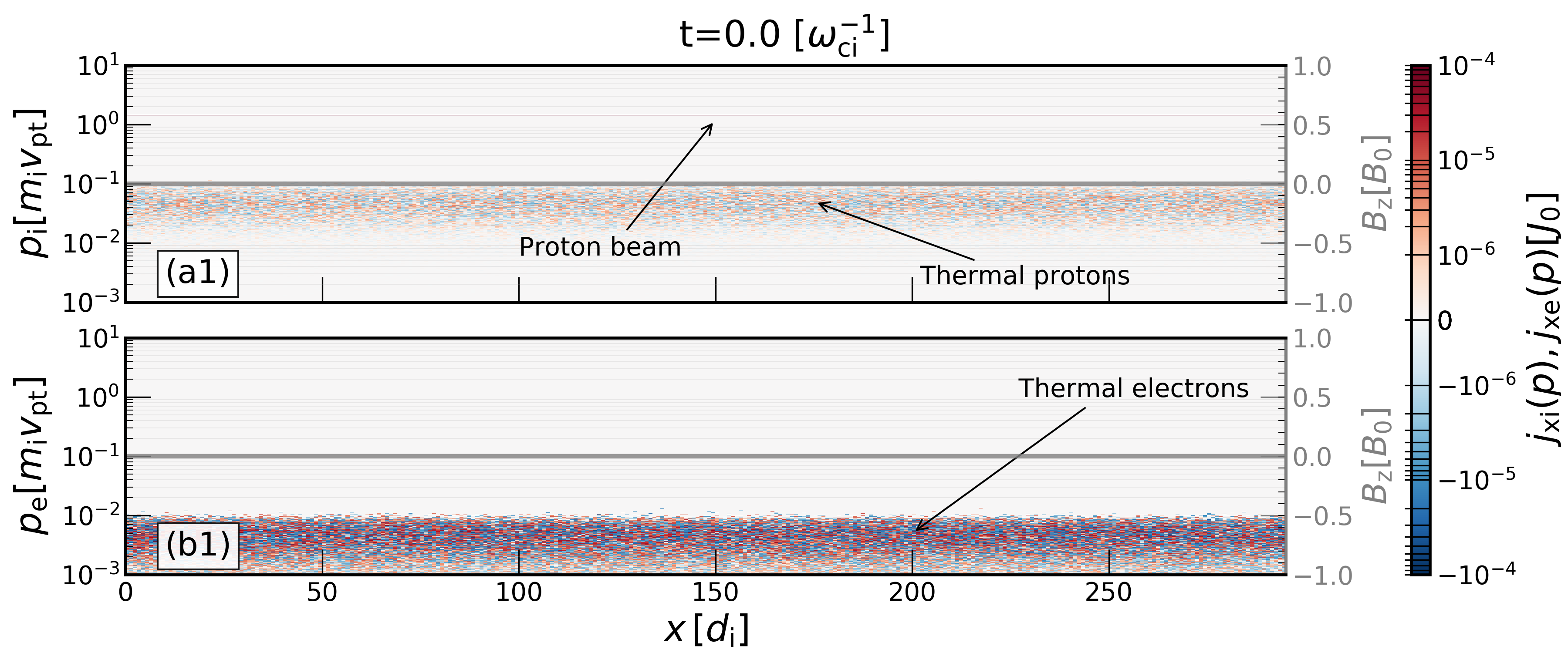}
\includegraphics[width=3.4in]{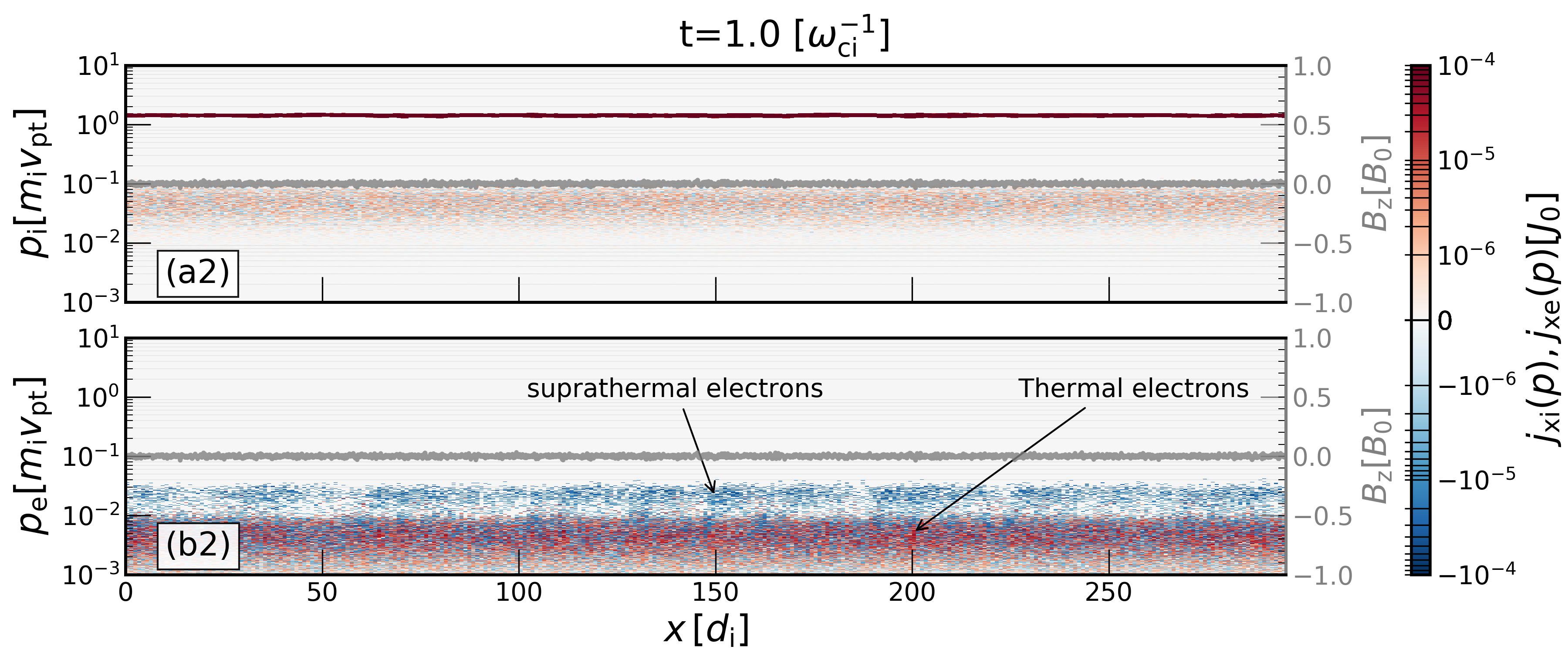}
\includegraphics[width=3.4in]{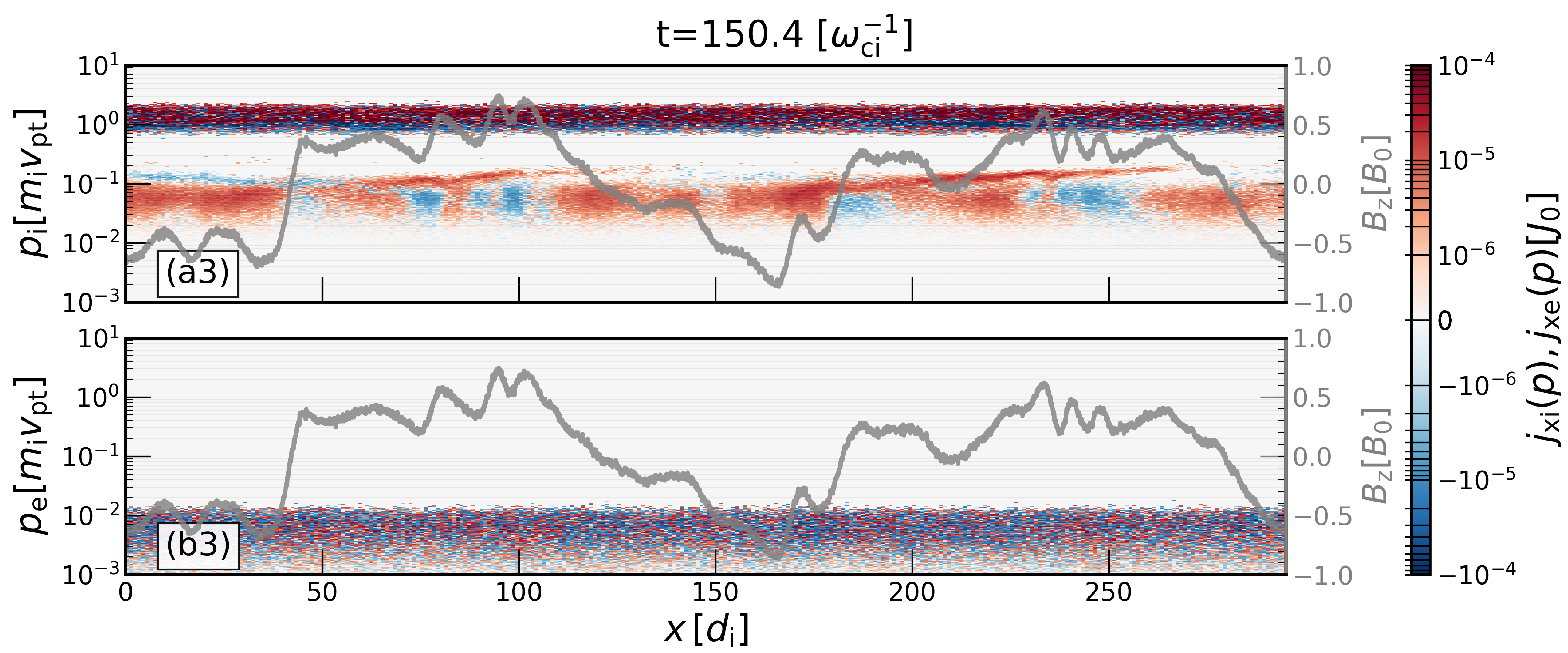}
\includegraphics[width=3.25in]{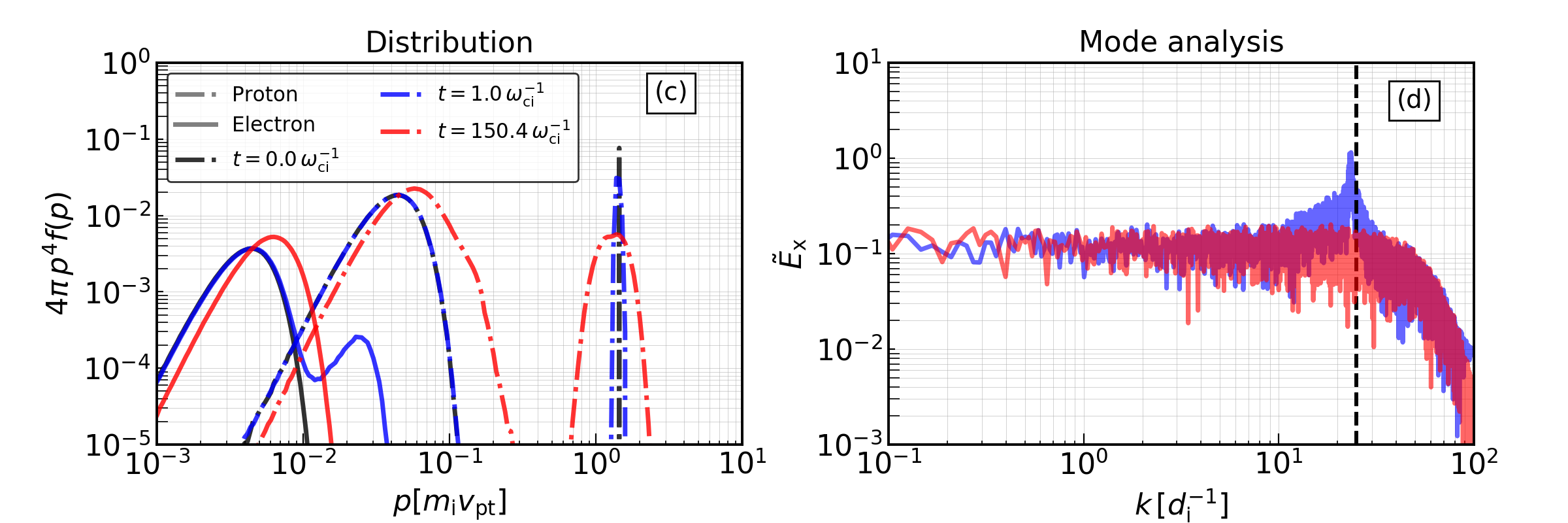}
\caption{A controlled simulation showing the development of return-current electrons due to a cold proton beam (region 1 in Table \ref{tab:controlsim}).
Panels (a1,a2,a3) and (b1,b2,b3) represent the $x-p$ phase-space of $j_{\rm x}$ for protons and electrons, along with profiles of the magnetic field (grey curves and right axes) at $t=0,\, 1,$ and $150\omega_{\rm ci}^{-1}$.
The corresponding spectra of protons and electrons and the Fourier analysis of $E_{\rm x}$ are displayed in panels (c) and (d), respectively.
The proton spectra also contain the beam distribution ($p\gtrsim m_{\rm i}v_{\rm pt}$, where $v_{\rm pt}=v_{\rm ib}/2$), which is re-normalized to keep the beam spectra within the chosen y-range.
In panel (d), the vertical line shows the predicted ES mode for the parameters used in this setup.
Development of ST electrons is evident at $t= 1 \omega_{\rm ci}^{-1}$.
In the late times, the ES mode and the ST hump in the electron distribution disappear and the whole background plasma attains a nonzero bulk motion along the direction of the beam.}
\label{fig:csimu_timesnap_1}
\end{figure}

We start with the case of the dilute beam with $n_{\rm ib}/n_{\rm 0}=0.005$.
Figures \ref{fig:csimu_timesnap_1}(a1) -- \ref{fig:csimu_timesnap_1}(b3) show the $x-p$ phase-space distribution of the current density $j_{\rm x}$ at $t=0,1,$ and $150\,\omega_{\rm ci}^{-1}$;
the grey lines indicate the profile of $B_{\rm z}$ (values on the right axes).
At $t=0$, i.e., Figures \ref{fig:csimu_timesnap_1}(a1) and \ref{fig:csimu_timesnap_1}(b1) shows that the current density is non-zero only for the proton beam ($j_{\rm x}\simeq 0.15\,e\, n_{\rm 0}\, v_{\rm A}$), while the background plasma has random fluctuations that average to zero.
Figures \ref{fig:csimu_timesnap_1}(c) and \ref{fig:csimu_timesnap_1}(d) show the spectra and the beam-induced modes respectively as discussed below.

At $t=1\,\omega_{\rm ci}^{-1}$, we observe the appearance of the ES mode and of ST electrons, as shown in Figures \ref{fig:csimu_timesnap_1}(b2), \ref{fig:csimu_timesnap_1}(c), and \ref{fig:csimu_timesnap_1}(d), similar to what happens far upstream (Figure \ref{fig:tracrealparticle}).
Moreover, for these parameters, the ratio of the beam momentum ($P_{\rm ib}=n_{\rm ib}p_{\rm ib}$) to the magnetic pressure $P_{\rm ib}/ P_{\rm B0} \approx 10$ satisfies the nonresonant instability \citep[e.g.,][especially equation 9]{bell04,gupta+21}, i.e., $\delta B/B_{\rm 0} \gtrsim 1$ can be expected \citep{gupta+21p}.
However, since the growth rate of the fastest-growing mode is $\gamma_{\rm fast}=0.5 (n_{\rm ib}/n_{\rm 0})(v_{\rm bi}/v_{\rm A}) \omega_{\rm ci} =0.08 \,\omega_{\rm ci}$, it takes a few ion cyclotron times to grow $B_{\rm z}$ appreciably.

At a later time, $t=150\,\omega_{\rm ci}^{-1}$, we see that the overall electron distribution has shifted to a high-temperature thermal-like distribution and the ES mode has faded away (Figures \ref{fig:csimu_timesnap_1}(c) and \ref{fig:csimu_timesnap_1}(d)).
Thus, the ST electrons that we have observed in this dilute cold beam simulation as well as in the far upstream can be considered the early phases of electron energization,
which is initially driven by ES instability and later by the EM turbulence of the streaming instabilities.
Eventually, the current is disrupted and the whole background moves along the direction of the beam, due to the linear momentum conservation.
Finally, we note that at saturation $\delta B/B_0\approx 1$, this is consistent with the fact that the Bell instability saturates at $B_{\rm z}/B_{\rm 0}\approx \sqrt{P_{\rm ib}/ P_{\rm B0}}\sim 1$, where $P_{\rm cr}/P_{\rm B0} \simeq 10$ \citep[e.g.,][]{gupta+21,gupta+21p,zacharegkas+22}.

When the beam is dense (representing region 3 in Table \ref{tab:controlsim}), we find a similar evolution, 
with the exceptions that: 1) the beam current $j_{\rm x}\simeq 3\, e\,n_{\rm 0}\,v_{\rm A}$ is too strong to drive nonresonant instability, since it makes $\gamma_{\rm fast}/\omega_{\rm ci}\equiv j_{\rm x}/(2\,n_{\rm 0}\,e\,v_{\rm A})> 1$, which does not favor the traditional nonresonant instability \citep[e.g.,][ Lichko et al, in prep]{zweibel+10}, and 2) the thermalization is achieved within $\lesssim 2\,\omega_{\rm ci}^{-1}$ (i.e., much earlier than that found in the dilute beam case), which we will elaborate in \S \ref{sec:eheating}.
%\dam{mention that it's the MTSI and announce our work with Emily.}
%===================================================
\subsubsection{Hot Beams}\label{subsec:hot}
%
%===================================================
In the shock precursor, the particles are diffusing and thus more isotropic, i.e., the beam cannot be considered cold.
To study the response of background electrons in this region, we use a hot beam ($u\gtrsim v_{\rm ib}$, see region $2$ in Table \ref{tab:controlsim}).
In this case, instead of a distinct $E_{\rm x}$ mode, broadband fluctuations arise similar to Figure \ref{fig:timesnap_fld}(b2) at $t=282.7\, \omega_{\rm ci}^{-1}$, i.e., when the current is mostly driven by diffusing particles.
In this scenario the response of the background plasma is incoherent on spatial scales of the $E_{\rm x}$ modes, which are approximately on the order of $2\pi (v_{\rm ib}/c) d_{\rm e}$ (Equation \ref{eq:kfast}).
As in the cold case, the later evolution is characterized by the growth of nonresonant modes and acceleration of the background plasma in the direction of the beam.

To summarize, our controlled simulations of different types of beams, corresponding to three distinct regions of the upstream (Table \ref{tab:controlsim}), show that current compensation invariably leads to the production of ST electrons, similar to those appearing in global shock simulation (\S \ref{sec:mechanism}).
We have also investigated the long-term evolution of the ST electrons and found that it becomes a thermal-like distribution.
While back-streaming ST electrons are indeed a transient phenomenon, they are always present far upstream, and may be mistakenly taken for accelerated particles. 
We have also investigated the long-term evolution of the ST electrons and found that ST electrons are thermalized, which contributes to overall heating, as shown in Figure \ref{fig:csimu_timesnap_1}(c).
Since the current is relatively small at large distances, we do not expect a substantial level of heating when averaging over all electrons in the far upstream region, and in general not beyond the heating that the Bell instability provides \citep[e.g.,][]{caprioli+14a,gupta+21p}.
However, we will show that this changes closer to the shock under the action of the dense beam of specularly-reflected protons in the shock foot (Figure \ref{fig:timesnap_x_vx}).
%%%%%%%%%%%%%%%%%%%%%%%%%%%%%%%%%%%%%%%%%%%%%%%%%%%%%%%%%%%
%%%%%%%%%%%%%%%%%%%%%%%%%%%%%%%%%%%%%%%%%%%%%%%%%%%%%%%%%%%
%%%%%%%%%%%%%%%%%%%%%%%  Section 2  %%%%%%%%%%%%%%%%%%%%%%%
%%%%%%%%%%%%%%%%%%%%%%%%%%%%%%%%%%%%%%%%%%%%%%%%%%%%%%%%%%%
%%%%%%%%%%%%%%%%%%%%%%%%%%%%%%%%%%%%%%%%%%%%%%%%%%%%%%%%%%%
\section{Electron heating in the shock foot} \label{sec:eheating}
%
%%%%%%%%%%%%%%%%%%%%%%%%%%%%%%%%%%%%%%%%%%%%%%%%%%%%%%%%%%%
\begin{figure}[b!]
\centering
\includegraphics[width=3.2in]{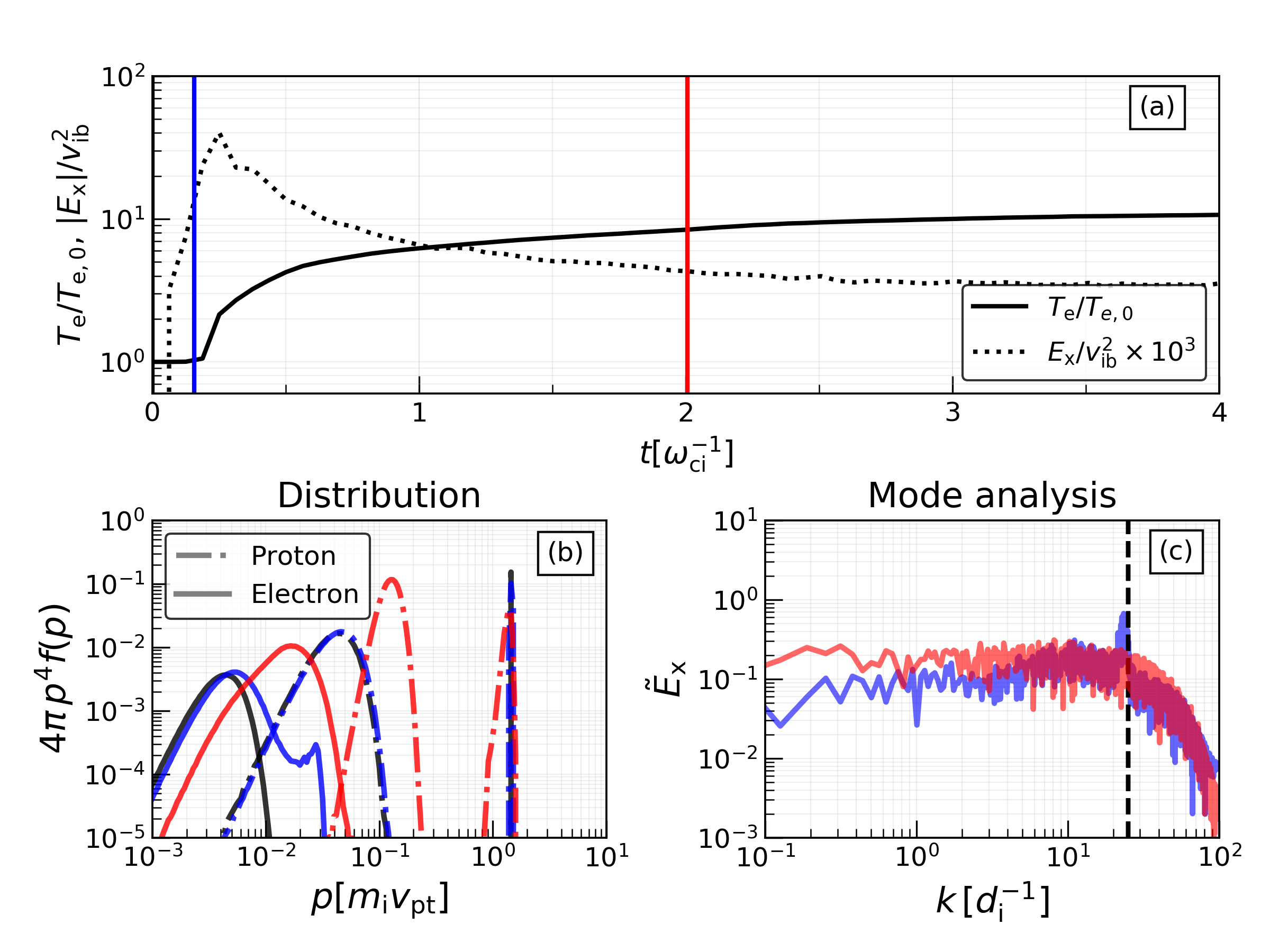}
\caption{
Response of background plasma in the case of cold dense beam (region 3 in Table \ref{tab:controlsim}).
Panel (a) displays the time evolution of $E_{\rm x}$ and $T_{\rm e}$.
Panels (b) and (c) represent the distribution of electrons and protons at two epochs marked in panel (a).
Note the net heating is more intense than that is found for the far upstream dilute beam case (Figure \ref{fig:csimu_timesnap_1}(c)).
}
\label{fig:dense_cold}
\end{figure}
Before encountering the shock, the upstream plasma has to experience the dense ($n_{\rm ib}/n_{\rm 0}\sim 0.1$) reflected proton beam in the foot.
Such an interaction is important for electron heating and, potentially, injection into DSA.
To quantify electron heating in the shock foot, we have performed a set of controlled simulations for a wide range of beam and background plasma parameters.
For each simulation, we calculate the effective electron temperature tensor  
\begin{equation} \label{eq:tmp}
%k_{\rm B}T_{\rm rs,e} \equiv \frac{1}{N_{\rm e}}\sum_{k=1}^{N_{\rm e}} \gamma_{\rm e}^{k}m_{\rm e}\,(u_{\rm r,e }^{\rm k}-v_{\rm r,e})(u_{\rm  s,e}^{\rm k}-v_{\rm  s,e})\,,
k_{\rm B} T_{\rm rs} \equiv \frac{1}{N_{\rm e}}\int \gamma\, u_{\rm r}(p')\, u_{\rm s}(p')\, f(p') d^3p'
\end{equation}
where $r,s\in x,y,z$, $N_{\rm e}$ is the total number of electrons in the computational domain, and $u_{\rm r}$ is the velocity in the $r$-direction, which is obtained by transforming it to the comoving plasma frame.
We call this an effective temperature because the electron distribution is not Maxwellian, and its shape evolves with time.
To find a correlation with the ES mode, we also estimate the volume averaged absolute value of $E_{\rm x}$.

Figure \ref{fig:dense_cold} presents the results of our benchmark dense cold beam simulation (region 3 in Table 3), where $m_{\rm R}=100$, $v_{\rm A}/c=1.33\times 10^{-2}$, $v_{\rm th}/c=6.67\times 10^{-3}$ (corresponding to a Debye length $=0.067\,d_{\rm e}$), $\Delta = d_{\rm e}/10$, $v_{\rm ib}/c=0.4$, and $n_{\rm ib}/n_0=0.1$.
Panel \ref{fig:dense_cold}(a) shows that there is an initial stage ($\sim 0.1\,\omega_{\rm ci}^{-1}$) when the effective temperature $T_{\rm rr}$ (hereafter, $T_{\rm e}$) does not change much, whereas $E_{\rm x}$ raises rapidly.
After $E_{\rm x}$ reaches a peak ($E_{\rm x}/v_{\rm pt}^2\sim 0.03$ [$\tilde B_{\rm 0}/c^2$]), it damps and $T_{\rm e}$ increases. 
When $t\gtrsim 1\,\omega_{\rm ci}^{-1}$, $T_{\rm e}$ tends to $\sim 9 T_{\rm e,0}$ ($T_{\rm e,0}$ being the initial temperature).
These initial and final stages are displayed in Figures \ref{fig:dense_cold}(b) and \ref{fig:dense_cold}(c), which show that 1) the ST hump evolves to a thermal-like distribution and 2) the $E_{\rm x}$ mode disappears, similar to the case of a dilute cold beam case (Figure \ref{fig:csimu_timesnap_1}).

\begin{figure}[t!]
\centering
\includegraphics[width=3.in]{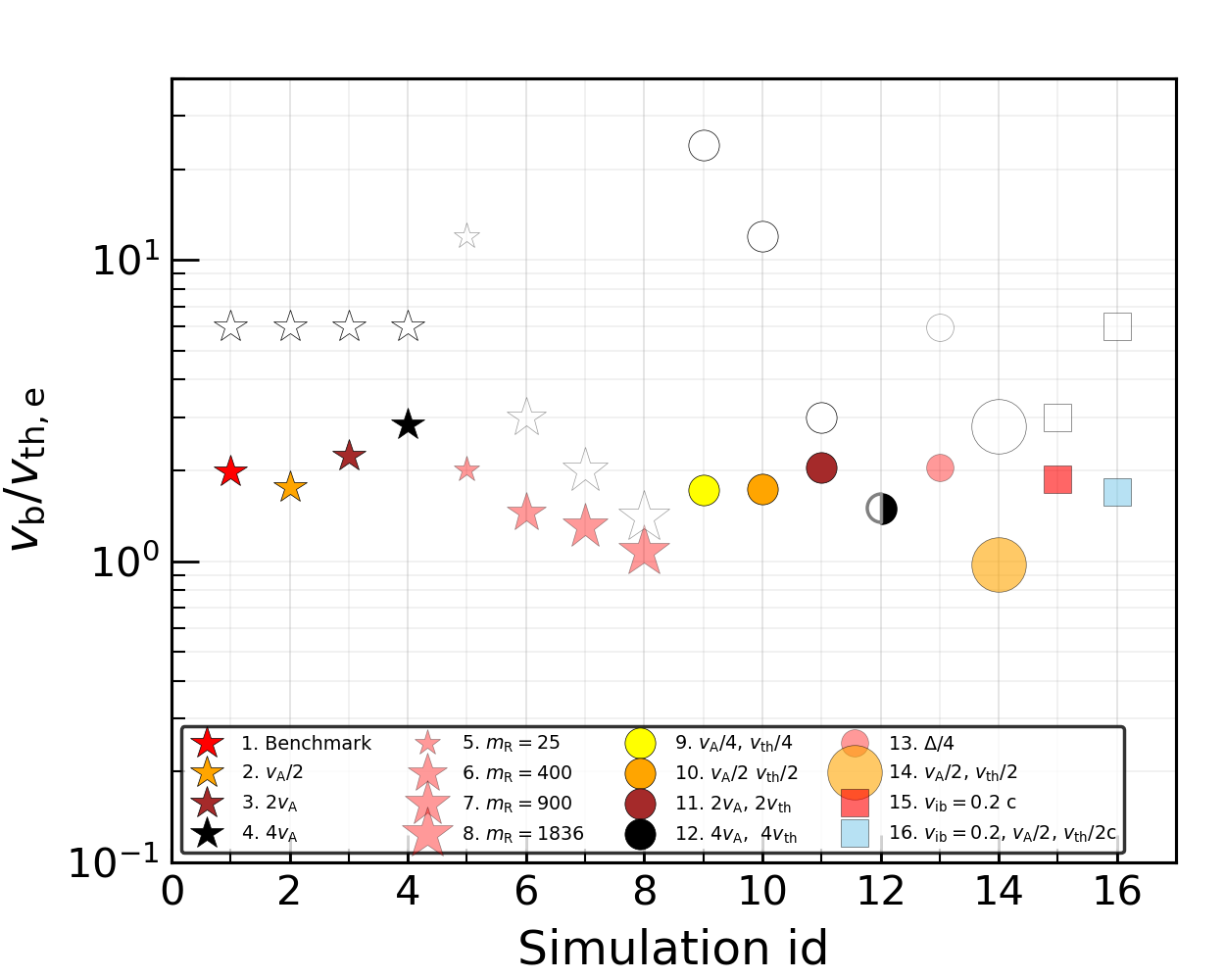}
\caption{
Level of electron heating for different parameters 
(mass-ratio $m_{\rm R}$, magnetization $v_{\rm A}$, thermal speed $v_{\rm th}$, beam velocity $v_{\rm ib}$, and the grid spacing $\Delta$).
The vertical axis displays the ratio of beam velocity ($v_{\rm ib}$) to the effective thermal speed of electrons $v_{\rm th,e}=(k_{\rm B} T_{\rm e}/m_{\rm e})^{1/2}$ obtained from a set of controlled simulations by varying one parameter at a time relative to the benchmark run, 
where the empty and filled symbols represent $t=0$ and the filled symbols $t= 2\, \omega_{\rm ci}^{-1}$, respectively. 
The horizontal axis denotes the id of different simulations (identical to the legend number). 
}
\label{fig:vtheff}
%\end{figure}
%\begin{figure}[t!]
\centering
\includegraphics[width=3in]{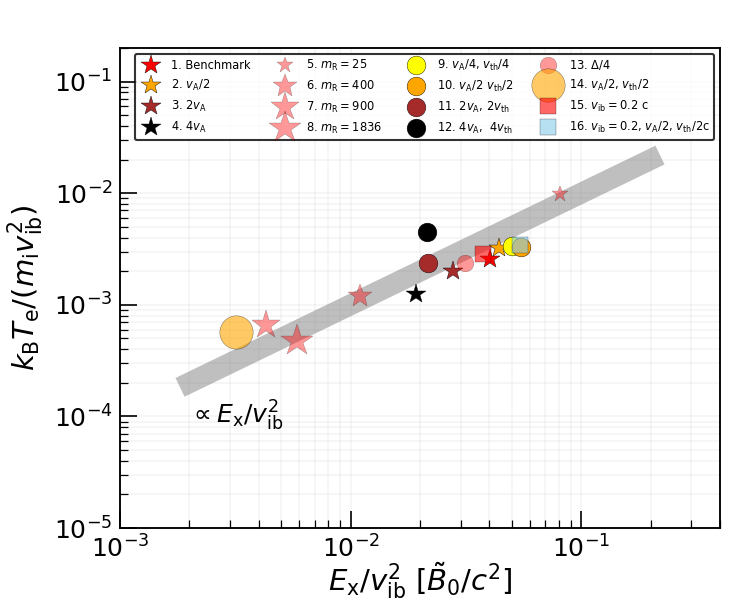}
\caption{
Correlation between the effective temperature of electrons at $t= 2\, \omega_{\rm ci}^{-1}$ and the peak amplitude of the beam-induced ES mode $E_{\rm x}/v_{\rm ib}^2$,
showing $T_{\rm e}\propto E_{\rm x}$.
}
\label{fig:Te_eff}
\end{figure}

The dependence of electron heating from the shock/simulation parameters is further detailed in Figures \ref{fig:vtheff} and \ref{fig:Te_eff}.
In both figures, the red star represents our benchmark dense cold beam simulation (discussed above), with other runs differing from the benchmark in one of these parameters, as listed in the legend.
We consider the evolution of the electron temperature from $t=0$ to $t=2\,\omega_{\rm ci}^{-1}$, physically corresponding to the time that it takes for the shock to overtake the shock foot, and compare the final electron thermal speed with the initial proton beam velocity;
note that $T_e$ is already very close to saturation, as shown in Figure \ref{fig:dense_cold}(a). 

For the benchmark run, Figure \ref{fig:vtheff} shows that the initial ratio $v_{\rm ib}/v_{\rm th,e}\simeq 6$ drops to $\approx 2$ (white and red stars, respectively), corresponding to an increase in electron temperature of a factor of $9$.
Very interestingly, changing the beam/background parameters does not change the fact that at saturation electrons are heated up until their thermal speed becomes of the order of the proton beam velocity, as shown in Figure \ref{fig:vtheff}. 
This universality can be explained by considering the beam-induced ES mode, which is the main driver of electron heating at the shock foot (see, e.g., Figure \ref{fig:csimu_timesnap_1}).

Figure \ref{fig:Te_eff} shows the peak amplitude of $E_{\rm x}$  for the simulations in Figure \ref{fig:vtheff}.
First, we observe a positive correlation between $T_{\rm e}$ and $E_{\rm x}$, as expected from the above discussion of the benchmark run (Figure \ref{fig:dense_cold}).
Second, we see that $E_{\rm x}\propto v_{\rm ib}^2$ (light-blue square versus the benchmark run) for a fixed beam density, which suggests that the strength of the ES mode depends on the kinetic energy density of the beam.
Third, $E_{\rm x}$ decreases with the mass ratio (red stars with increasing size), yet yielding the same final $v_{\rm ib}/v_{\rm th,e}$.
This trend can be accounted for by noticing that a larger $m_{\rm R}$ reduces the difference between $v_{\rm th,e}$ and $v_{\rm ib}$ that enters the dispersion relations (see Appendix \ref{app:dispersion}).
This suggests that for realistic values of $m_{\rm R}$ the amplitude of the current-induced $E_{\rm x}$ mode may decrease, but without changing the final $v_{\rm ib}/v_{\rm th,e}$ ratio.
We also verified that the final temperature is insensitive to the grid resolution (red star vs red circle).

Since in general $v_{\rm ib}\sim v_{\rm sh}$, in the absence of upstream heating the electron sonic Mach number $\mathcal{M}_{\rm s,e,0} \equiv v_{\rm sh}/v_{\rm th,e}$ could span orders of magnitude, from $\mathcal{M}_{\rm s,e}\sim 1$ in heliospheric and intracluster shocks to $\mathcal{M}_{\rm s,e}\gg 1$ shocks in the interstellar medium.
However, we have shown that, when the proton beam velocity is larger than the thermal speed of background electrons, the shock foot is prone to an ES instability driven by an initial current imbalance, which grows on a time scale $\ll \omega_{\rm ci}^{-1}$.
The energy associated with the fluctuating $E_{\rm x}$ is quickly transferred to the background, and in particular to the electrons, which heat up to $v_{\rm th,e}\approx v_{\rm ib}$. 
As a result, the effective sonic Mach number for electrons in the shock foot becomes $M_{\rm s,e} \equiv v_{\rm ib}/v_{\rm th,e}\approx 1-3 $ (see Figure \ref{fig:vtheff}), independent of the initial sonic Mach number.
While it may be a semantic question whether heating in the foot can be considered to occur \emph{upstream} of the shock versus at the shock, this finding has implications not only for the temperature that electrons achieve behind the shock, but also for the injection of electrons into DSA.
In fact, a larger thermal speed facilitates magnetic mirroring \citep[e.g.,][]{guo+14a} and hence electron reflection and acceleration \citep[][]{gupta+23c}.
%%%%%%%%%%%%%%%%%%%%%%%%%%%%%%%%%%%%%%%%%%%%%%%%%%%%%%%%%%%
%%%%%%%%%%%%%%%%%%%%%%%%%%%%%%%%%%%%%%%%%%%%%%%%%%%%%%%%%%%
%%%%%%%%%%%%%%%%%%%%%%%  Section 7  %%%%%%%%%%%%%%%%%%%%%%%
%%%%%%%%%%%%%%%%%%%%%%%%%%%%%%%%%%%%%%%%%%%%%%%%%%%%%%%%%%%
%%%%%%%%%%%%%%%%%%%%%%%%%%%%%%%%%%%%%%%%%%%%%%%%%%%%%%%%%%%
\section{Return currents in Quasi-perpendicular Shocks}\label{sec:discuss}
%
%%%%%%%%%%%%%%%%%%%%%%%%%%%%%%%%%%%%%%%%%%%%%%%%%%%%%%%%%%%
\begin{figure}
\centering
\includegraphics[width=3.4in]{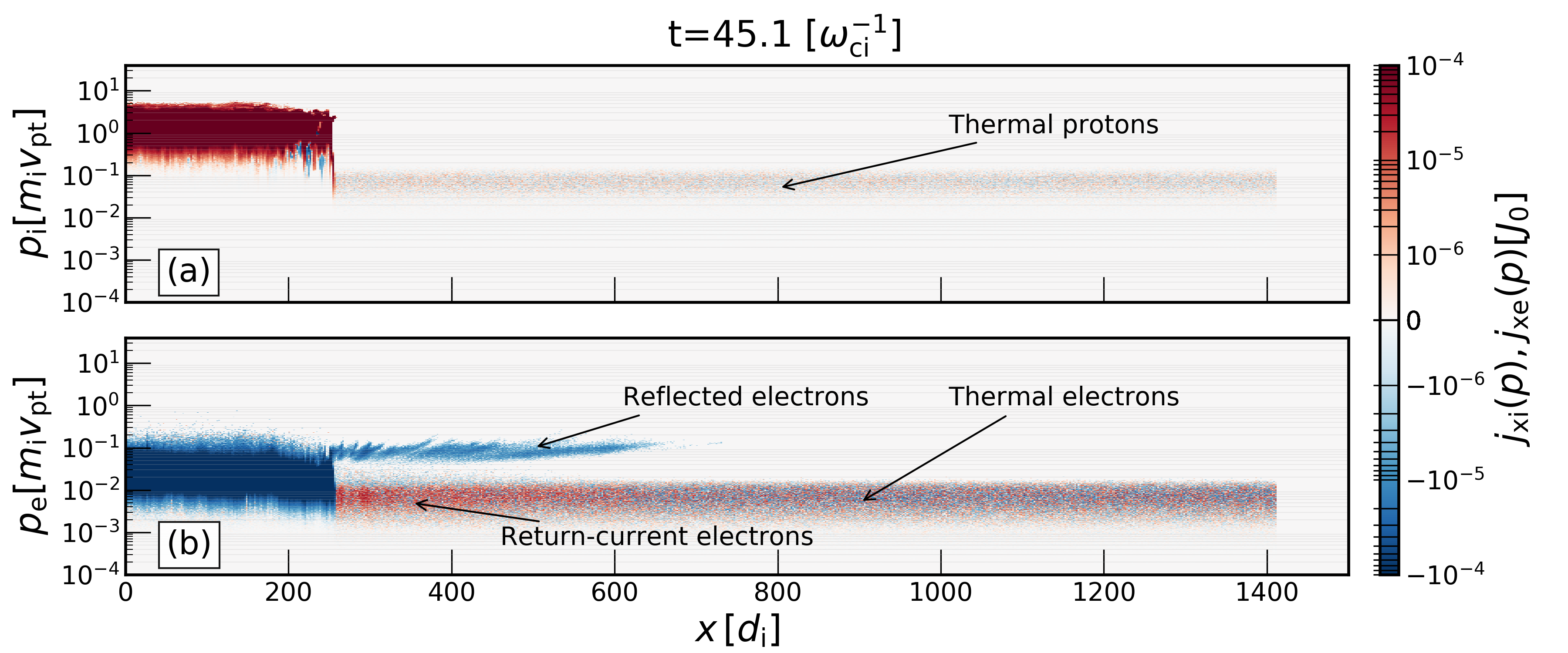}
\caption{
Current diagnostics for a quasi-perpendicular ($\theta_{\rm Bn}=63^{\rm o}$) shock.
The panels (a) and (b) stand for protons and electrons, respectively.
The distinct populations upstream are marked by the arrows (also compare with Figure \ref{fig:keyfig}(f1,f2)).
}
\label{fig:jx_in_qperp}
\end{figure}
So far we have discussed return currents in the context of quasi-parallel shocks, but it is worth discussing what happens when the shock is quasi-perpendicular, too.
It is already known that oblique shocks reflect more electrons than protons \citep[][]{guo+14a,guo+14b,xu+20,morris+22,bohdan+22}, which implies that the unbalanced current in the upstream should be carried by back-streaming electrons and that the return current must be positive.

We perform simulations of quasi-perpendicular shocks with $\theta_{\rm Bn}=63\deg$, keeping all the other parameters identical to our benchmark shock run ($v_{\rm pt}=0.2\, c, \mathcal{M}_{\rm A}=20, \mathcal{M}_{\rm s}=40$, and $m_{\rm R}=100$).
Figures \ref{fig:jx_in_qperp}(a) and \ref{fig:jx_in_qperp}(b) show the phase-space distribution of current density ($j_{\rm x}$) for protons and electrons at $t=45\,\omega^{-1}_{\rm ci}$, respectively.
Figure \ref{fig:jx_in_qperp}(a) shows that the proton population upstream is completely thermal, while the electron population in Figure \ref{fig:jx_in_qperp}(b) displays three district features, as labeled.
In this scenario, the unbalanced current is induced by the reflected electrons, which also develop the ES mode \citep[][]{bohdan+22}.
The return current is carried by background electrons, which move toward the shock, while staying close to thermal.
In fact, the current in back-streaming electrons does not carry much momentum/energy, thereby not leading to any appreciable heating.

%Our study demonstrates that any shock, regardless of its inclination, produces currents that are balanced by the mobile thermal electrons;
%such electrons may become suprathermal (if the driving current is strong enough) and either move toward the shock or upstream, depending on the direction of the dominant current in the driving beam.

Our study demonstrates that the distributions of upstream particles readjust automatically to neutralize the total currents in the plasma, regardless of the initial inclination of the magnetic field relative to the shock normal. 
These processes primarily affect electrons: if the driving current carries strong enough momentum, then electrons may become hot/supra-thermal and either move toward the shock or upstream, depending on the direction of the dominant current in the driving beam.
While throughout this work we take $v_{\rm pt}/c=0.2$ to discuss the production mechanism and the evolution of supra-thermal electrons, we find that the results persist even when considering smaller values of $v_{\rm pt}/c$, in particular when the upstream electrons are initially cold and moderately less magnetized, as typically found in a high Mach number shock ($\mathcal{M}_{\rm s,e}> 1$ and $\mathcal{M}_{\rm A}\gtrsim 10$).
We expect that the general conclusions should qualitatively hold for $v_{\rm pt}/c > 0.2$. 
An in-depth examination of trans-relativistic shocks is beyond the scope of present work, since multi-dimensional effects are crucial for the self-generation of NT current beam upstream of such shocks \citep[][]{crumley+19}.
%%%%%%%%%%%%%%%%%%%%%%%%%%%%%%%%%%%%%%%%%%%%%%%%%%%%%%%%%%%
%%%%%%%%%%%%%%%%%%%%%%%%%%%%%%%%%%%%%%%%%%%%%%%%%%%%%%%%%%%
%%%%%%%%%%%%%%%%%%%%%%%  Section 7  %%%%%%%%%%%%%%%%%%%%%%%
%%%%%%%%%%%%%%%%%%%%%%%%%%%%%%%%%%%%%%%%%%%%%%%%%%%%%%%%%%%
%%%%%%%%%%%%%%%%%%%%%%%%%%%%%%%%%%%%%%%%%%%%%%%%%%%%%%%%%%%
\section{Summary}\label{sec:conclusion}
In this work we have used particle-in-cell simulations to study the balance of electric currents upstream of collisionless shocks and their impact on electron distribution.
We have developed a diagnostic to identify the return-current populations and confirmed our results by designing a controlled test problem.
Our main findings are summarized below:
\begin{enumerate}
\setlength\itemsep{-0.005\textwidth}
    \item Whenever an unbalanced current is produced in the plasma due to particles reflected from the shock, the system pulls electrons out of the thermal background via an electrostatic process akin to the two-stream instability (Figure \ref{fig:csimu_timesnap_1}).
    \item Depending on the direction of the unbalanced current in the upstream frame, compensating electrons can either move toward the upstream infinity or toward the shock (back-streaming or forward-streaming electrons, respectively).
    The former scenario is typical of quasi-parallel shocks, where the current is carried by nonthermal (NT) ions (Figure \ref{fig:keyfig}).
    The latter scenario occurs for quasi-perpendicular shocks, where the currents are typically made of electrons (Figure \ref{fig:jx_in_qperp}).
    \item In the shock upstream, the driving current-induced electrostatic mode transfers its energy to background thermal electrons and generates a population of supra-thermal (ST) electrons. 
    Such ST electrons carry the return current and can be found anywhere from the far upstream to the shock foot (Figures \ref{fig:tracrealparticle} and \ref{fig:timesnap_x_vx}).
    Given the non-thermal nature of these electrons, they may be easily confused with electrons that are experiencing DSA.
    We show that back-streaming ST electrons can be produced locally in the upstream well before they interact with the shock (Figure \ref{fig:tracrealparticle}). 
    \item During the initial stages of interaction between the current beam and the thermal plasma (either far upstream or during the quasi-periodic shock reformation), the distribution of the return current electrons may appear as a ST hump attached to the thermal distribution.
    Then, while being advected towards the shock, these electrons are progressively thermalized (Figure \ref{fig:dense_cold})
     \item In the shock foot there periodically appears a strong unbalanced proton current due to the shock reformation (Figure \ref{fig:timesnap_x_vx}).   
     A crucial byproduct of such a current imbalance in the shock foot is the strong heating of thermal electrons.
     This effect is noticeable mainly when the thermal speed of background electrons is smaller than the velocity of the current-producing beam (\S \ref{sec:eheating}).
     By performing a set of controlled simulations, we find that the final thermal speed of the electrons generally becomes comparable to the velocity of the proton beam (Figure \ref{fig:vtheff}).
    This is a non-trivial finding which suggests that the effective electron sonic Mach number close to the shock should generally be $\lesssim 1-3$. This energization can potentially simplify the process of reflecting and injecting electrons into DSA.
\end{enumerate}

To conclude, we have detailed the production mechanism and the properties of the electrons that make up return currents in collisionless shocks.
How these results may help fostering our understanding of injection and DSA of electrons will be discussed in forthcoming papers.

\section*{ACKNOWLEDGMENTS}
SG thanks Emily Lichko and Vladimir Zekovic for insightful discussions.
We acknowledge the computational resources provided by the University of Chicago Research Computing Center and ACCESS (TG-AST180008). 
DC was partially supported by NASA (grants 80NSSC18K1218, 80NSSC20K1273, and 80NSSC18K1726) and by NSF (grants AST-1714658, AST-2009326, AST-1909778, PHY-1748958, and PHY-2010240). 
AS acknowledges the support of NSF grants PHY-2206607 and AST-1814708.

\appendix
%%%%%%%%%%%%%%%%%%%%%%%%%%%%%%%%%%%%%%%%%%%%%%%%%%%%
\section{A cold current beam in a warm electron-ion plasma}\label{app:dispersion}
%
%%%%%%%%%%%%%%%%%%%%%%%%%%%%%%%%%%%%%%%%%%%%%%%%%%%%
To understand the role of electron temperature in a beam-plasma system, we investigate its dispersion equation.
Considering that the system consists of three species respectively background protons, background electrons, and a cold beam made of either protons or electrons, we obtain a dispersion equation for the modes parallel to the beam direction:
\begin{equation}\label{eq:dispersion1}
    1 \simeq \frac{\omega_{\rm pi}^2}{\omega^2} + \frac{\omega_{\rm pe}^2}{\omega^2- a_{\rm th,e}^2\,k^2} + \frac{n_{\rm b}/n_{\rm 0} \omega_{\rm b}^2/\gamma_{\rm b}^3}{(\omega- v_{\rm b}\,k)^2}.
\end{equation}
Here we have assumed that the electrons have non-negligible sound speed $a_{\rm th,e}(\equiv \sqrt{5/3}v_{\rm th,e})$, the magnetic field is negligible, the beam density $n_{\rm b}$, and the Lorentz factor $\gamma_{\rm b}$.
A similar equation can be found in \citet{bret+10b}, where the system consists of a beam of relativistic electrons and a cold background plasma.

\begin{figure}
    \centering
    \includegraphics[width=2.15in]{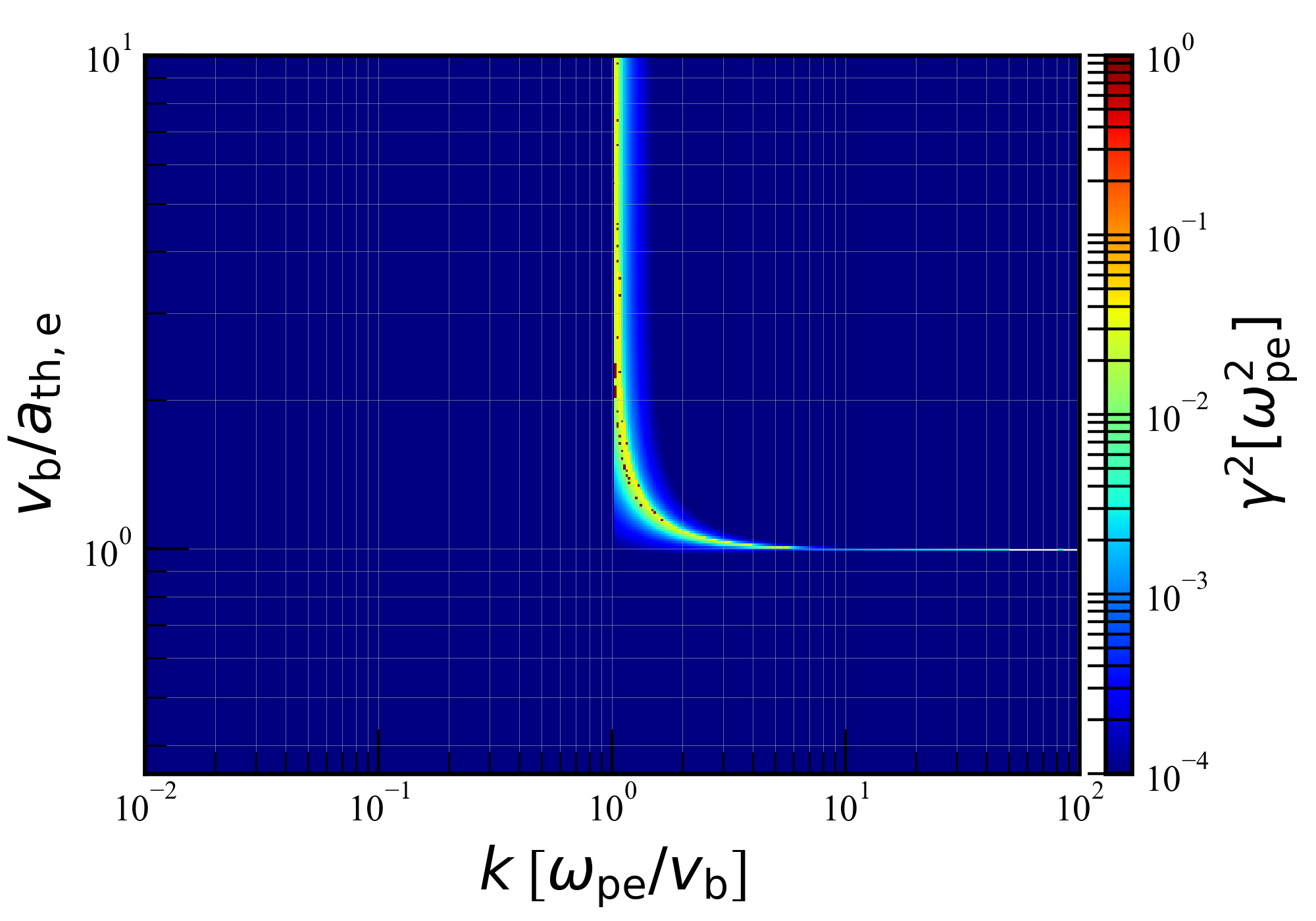}
    \caption{Solution of the dispersion Equation \ref{eq:dispersion1} as function of $k$ for different values of $v_{\rm b}/a_{\rm th,e}$ and a fixed $n_{\rm b}/n_{\rm 0}$.}
    \label{fig:dispersion}
\end{figure}

Substituting $\omega$ by $\omega_{\rm R}+ j\, \gamma$ (where $\omega_{\rm R}$ and $\gamma$ represent the real and growing/damping terms, and $j=\sqrt{-1}$), Equation \ref{eq:dispersion1} can be solved numerically.
To simplify, we have assumed that the background protons are undisturbed, i.e., the first term in the right-hand side of Equation \ref{eq:dispersion1} can be neglected, which is reasonable in the limit $m_{\rm R}\gg 1$.
We also consider a case where $\gamma_{\rm b}\rightarrow 1$ (a larger $\gamma_{\rm b}$ reduces the effective $n_{\rm b}/n_{\rm 0}$). 
 The solutions for $\gamma^2$ for different $k$ and $(v_{\rm b}/a_{\rm th,e})$ are displayed in Figure \ref{fig:dispersion}.
The figure indicates that the mode corresponding to the fastest growth rate is $k\approx \omega_{\rm pe}/v_{\rm b}$ (Equation \ref{eq:kfast}) and 
the growth-rate drops to zero, when $v_{\rm b}/a_{\rm th,e}<1 $. 
Therefore, whenever the thermal speed of electrons is smaller than the beam velocity, the contribution from the ES mode becomes significant, which raises the electron temperature to stabilize the system, as elaborated in section \ref{sec:eheating}.
\bibliography{Total}{}

\begin{thebibliography}{}
\expandafter\ifx\csname natexlab\endcsname\relax\def\natexlab#1{#1}\fi
\providecommand{\url}[1]{\href{#1}{#1}}
\providecommand{\dodoi}[1]{doi:~\href{http://doi.org/#1}{\nolinkurl{#1}}}
\providecommand{\doeprint}[1]{\href{http://ascl.net/#1}{\nolinkurl{http://ascl.net/#1}}}
\providecommand{\doarXiv}[1]{\href{https://arxiv.org/abs/#1}{\nolinkurl{https://arxiv.org/abs/#1}}}

\bibitem[{{Amato} \& {Blasi}(2009)}]{amato+09}
{Amato}, E., \& {Blasi}, P. 2009, MNRAS, 392, 1591,
  \dodoi{10.1111/j.1365-2966.2008.14200.x}

\bibitem[{{Arbutina} \& {Zekovi{\'c}}(2021)}]{arbutina+21}
{Arbutina}, B., \& {Zekovi{\'c}}, V. 2021, Astroparticle Physics, 127, 102546,
  \dodoi{10.1016/j.astropartphys.2020.102546}

\bibitem[{{Axford} {et~al.}(1978){Axford}, {Leer}, \& {Skadron}}]{axford+78}
{Axford}, W.~I., {Leer}, E., \& {Skadron}, G. 1978, in International Cosmic Ray
  Conference, Vol.~11, ICRC, 132--137.
\newblock \url{http://adsabs.harvard.edu/abs/1978ICRC...11..132A}

\bibitem[{{Bell}(1978)}]{bell78a}
{Bell}, A.~R. 1978, MNRAS, 182, 147.
\newblock \url{https://ui.adsabs.harvard.edu/abs/1978MNRAS.182..147B/abstract}

\bibitem[{{Bell}(2004)}]{bell04}
---. 2004, MNRAS, 353, 550, \dodoi{10.1111/j.1365-2966.2004.08097.x}

\bibitem[{{Bell} {et~al.}(2013){Bell}, {Schure}, {Reville}, \&
  {Giacinti}}]{bell+13}
{Bell}, A.~R., {Schure}, K.~M., {Reville}, B., \& {Giacinti}, G. 2013, MNRAS,
  431, 415, \dodoi{10.1093/mnras/stt179}

\bibitem[{{Blandford} \& {Ostriker}(1978)}]{blandford+78}
{Blandford}, R.~D., \& {Ostriker}, J.~P. 1978, ApJL, 221, L29,
  \dodoi{10.1086/182658}

\bibitem[{{Bohdan} {et~al.}(2019){Bohdan}, {Niemiec}, {Pohl}, {Matsumoto},
  {Amano}, \& {Hoshino}}]{bohdan+19a}
{Bohdan}, A., {Niemiec}, J., {Pohl}, M., {et~al.} 2019, \apj, 878, 5,
  \dodoi{10.3847/1538-4357/ab1b6d}

\bibitem[{{Bohdan} {et~al.}(2022){Bohdan}, {Weidl}, {Morris}, \&
  {Pohl}}]{bohdan+22}
{Bohdan}, A., {Weidl}, M.~S., {Morris}, P.~J., \& {Pohl}, M. 2022, Physics of
  Plasmas, 29, 052301, \dodoi{10.1063/5.0084544}

\bibitem[{{Bret} \& {Dieckmann}(2010)}]{bret+10b}
{Bret}, A., \& {Dieckmann}, M.~E. 2010, Physics of Plasmas, 17, 032109,
  \dodoi{10.1063/1.3357336}

\bibitem[{{Caprioli} {et~al.}(2020){Caprioli}, {Haggerty}, \&
  {Blasi}}]{caprioli+20}
{Caprioli}, D., {Haggerty}, C.~C., \& {Blasi}, P. 2020, \apj, 905, 2,
  \dodoi{10.3847/1538-4357/abbe05}

\bibitem[{{Caprioli} {et~al.}(2015){Caprioli}, {Pop}, \&
  {Spitkovsky}}]{caprioli+15}
{Caprioli}, D., {Pop}, A., \& {Spitkovsky}, A. 2015, \apjl, 798, 28.
\newblock \doarXiv{1409.8291}

\bibitem[{{Caprioli} \& {Spitkovsky}(2014{\natexlab{a}})}]{caprioli+14a}
{Caprioli}, D., \& {Spitkovsky}, A. 2014{\natexlab{a}}, \apj, 783, 91,
  \dodoi{10.1088/0004-637X/783/2/91}

\bibitem[{{Caprioli} \& {Spitkovsky}(2014{\natexlab{b}})}]{caprioli+14b}
---. 2014{\natexlab{b}}, \apj, 794, 46, \dodoi{10.1088/0004-637X/794/1/46}

\bibitem[{Crumley {et~al.}(2019)Crumley, Caprioli, Markoff, \&
  Spitkovsky}]{crumley+19}
Crumley, P., Caprioli, D., Markoff, S., \& Spitkovsky, A. 2019, \mnras, 485,
  5105, \dodoi{10.1093/mnras/stz232}

\bibitem[{{Giacalone} {et~al.}(1993){Giacalone}, {Burgess}, {Schwartz}, \&
  {Ellison}}]{giacalone+93}
{Giacalone}, J., {Burgess}, D., {Schwartz}, S.~J., \& {Ellison}, D.~C. 1993,
  \apj, 402, 550, \dodoi{10.1086/172157}

\bibitem[{Guo {et~al.}(2014{\natexlab{a}})Guo, Sironi, \& Narayan}]{guo+14a}
Guo, X., Sironi, L., \& Narayan, R. 2014{\natexlab{a}}, \apj, 794, 153,
  \dodoi{10.1088/0004-637X/794/2/153}

\bibitem[{Guo {et~al.}(2014{\natexlab{b}})Guo, Sironi, \& Narayan}]{guo+14b}
---. 2014{\natexlab{b}}, \apj, 797, 47, \dodoi{10.1088/0004-637X/797/1/47}

\bibitem[{{Gupta} {et~al.}(2022){Gupta}, {Caprioli}, \& {Haggerty}}]{gupta+21p}
{Gupta}, S., {Caprioli}, D., \& {Haggerty}, C. 2022, in 37th International
  Cosmic Ray Conference, 484, \dodoi{10.22323/1.395.0484}

\bibitem[{{Gupta} {et~al.}(2021){Gupta}, {Caprioli}, \& {Haggerty}}]{gupta+21}
{Gupta}, S., {Caprioli}, D., \& {Haggerty}, C.~C. 2021, \apj, 923, 208,
  \dodoi{10.3847/1538-4357/ac23cf}

\bibitem[{{Gupta} {et~al.}(in prep 2023){Gupta}, {Caprioli}, \&
  {Spitkovsky}}]{gupta+23c}
{Gupta}, S., {Caprioli}, D., \& {Spitkovsky}, A. in prep 2023, ApJ, C,
  \dodoi{123}

\bibitem[{{Haggerty} \& {Caprioli}(2020)}]{haggerty+20}
{Haggerty}, C.~C., \& {Caprioli}, D. 2020, \apj, 905, 1,
  \dodoi{10.3847/1538-4357/abbe06}

\bibitem[{{Hoshino} \& {Shimada}(2002)}]{hoshino+02}
{Hoshino}, M., \& {Shimada}, N. 2002, \apj, 572, 880, \dodoi{10.1086/340454}

\bibitem[{{Kumar} \& {Reville}(2021)}]{kumar+21}
{Kumar}, N., \& {Reville}, B. 2021, \apjl, 921, L14,
  \dodoi{10.3847/2041-8213/ac30e0}

\bibitem[{{Lee} {et~al.}(2004){Lee}, {Chapman}, \& {Dendy}}]{lee+04}
{Lee}, R.~E., {Chapman}, S.~C., \& {Dendy}, R.~O. 2004, \apj, 604, 187,
  \dodoi{10.1086/381881}

\bibitem[{{Morlino} {et~al.}(2010){Morlino}, {Amato}, {Blasi}, \&
  {Caprioli}}]{morlino+10}
{Morlino}, G., {Amato}, E., {Blasi}, P., \& {Caprioli}, D. 2010, \mnras, 405,
  L21, \dodoi{10.1111/j.1745-3933.2010.00851.x}

\bibitem[{{Morris} {et~al.}(2022){Morris}, {Bohdan}, {Weidl}, \&
  {Pohl}}]{morris+22}
{Morris}, P.~J., {Bohdan}, A., {Weidl}, M.~S., \& {Pohl}, M. 2022, \apj, 931,
  129, \dodoi{10.3847/1538-4357/ac69c7}

\bibitem[{{Muschietti} \& {Lemb{\`e}ge}(2017)}]{muschietti+17}
{Muschietti}, L., \& {Lemb{\`e}ge}, B. 2017, Annales Geophysicae, 35, 1093,
  \dodoi{10.5194/angeo-35-1093-2017}

\bibitem[{{Orusa} \& {Caprioli}(2023)}]{orusa+23}
{Orusa}, L., \& {Caprioli}, D. 2023, \prl, 131, 095201,
  \dodoi{10.1103/PhysRevLett.131.095201}

\bibitem[{{Park} {et~al.}(2015){Park}, {Caprioli}, \& {Spitkovsky}}]{park+15}
{Park}, J., {Caprioli}, D., \& {Spitkovsky}, A. 2015, Physical Review Letters,
  114, 085003, \dodoi{10.1103/PhysRevLett.114.085003}

\bibitem[{{Riquelme} \& {Spitkovsky}(2009)}]{riquelme+09}
{Riquelme}, M.~A., \& {Spitkovsky}, A. 2009, \apj, 694, 626,
  \dodoi{10.1088/0004-637X/694/1/626}

\bibitem[{{Shalaby} {et~al.}(2022){Shalaby}, {Lemmerz}, {Thomas}, \&
  {Pfrommer}}]{shalaby+22}
{Shalaby}, M., {Lemmerz}, R., {Thomas}, T., \& {Pfrommer}, C. 2022, \apj, 932,
  86, \dodoi{10.3847/1538-4357/ac6ce7}

\bibitem[{Sironi {et~al.}(2013)Sironi, Spitkovsky, \& Arons}]{sironi+13}
Sironi, L., Spitkovsky, A., \& Arons, J. 2013, \apj, 771, 54,
  \dodoi{10.1088/0004-637X/771/1/54}

\bibitem[{{Spitkovsky}(2005)}]{spitkovsky05}
{Spitkovsky}, A. 2005, in American Institute of Physics Conference Series, Vol.
  801, Astrophysical Sources of High Energy Particles and Radiation, ed.
  T.~{Bulik}, B.~{Rudak}, \& G.~{Madejski}, 345--350, \dodoi{10.1063/1.2141897}

\bibitem[{{Thomas} {et~al.}(1990){Thomas}, {Winske}, \& {Omidi}}]{thomas+90}
{Thomas}, V.~A., {Winske}, D., \& {Omidi}, N. 1990, \jgr, 95, 18809,
  \dodoi{10.1029/JA095iA11p18809}

\bibitem[{{Treumann}(2009)}]{treumann09}
{Treumann}, R.~A. 2009, \aapr, 17, 409, \dodoi{10.1007/s00159-009-0024-2}

\bibitem[{{Wilson} {et~al.}(2016){Wilson}, {Sibeck}, {Turner}, {Osmane},
  {Caprioli}, \& {Angelopoulos}}]{wilson+16}
{Wilson}, L.~B., {Sibeck}, D.~G., {Turner}, D.~L., {et~al.} 2016, \prl, 117,
  215101, \dodoi{10.1103/PhysRevLett.117.215101}

\bibitem[{{Xu} {et~al.}(2020){Xu}, {Spitkovsky}, \& {Caprioli}}]{xu+20}
{Xu}, R., {Spitkovsky}, A., \& {Caprioli}, D. 2020, \apjl, 897, L41,
  \dodoi{10.3847/2041-8213/aba11e}

\bibitem[{{Zacharegkas} {et~al.}(2022){Zacharegkas}, {Caprioli}, \&
  {Haggerty}}]{zacharegkas+22}
{Zacharegkas}, G., {Caprioli}, D., \& {Haggerty}, C. 2022, arXiv e-prints,
  arXiv:2210.08072.
\newblock \doarXiv{2210.08072}

\bibitem[{Zweibel \& Everett(2010)}]{zweibel+10}
Zweibel, E.~G., \& Everett, J.~E. 2010, The Astrophysical Journal, 709, 1412,
  \dodoi{10.1088/0004-637x/709/2/1412}

\end{thebibliography}
\bibliographystyle{aasjournal}
\end{document}